\documentclass[12pt]{article}

\input{epsf}
\usepackage{amsfonts}
\usepackage{amsmath}
\usepackage{amssymb}
\usepackage{latexsym}
\usepackage[dvips]{graphicx}
\usepackage{color}

\newcommand{\be}{\begin{equation}}
\newcommand{\ee}{\end{equation}}
\newcommand{\bea}{\begin{eqnarray}}
\newcommand{\eea}{\end{eqnarray}}
\newcommand{\beq}{\begin{equation}}
\newcommand{\eeq}{\end{equation}}

\numberwithin{equation}{section}

\begin{document}
\baselineskip 20pt
\begin{titlepage}
\begin{center}
{\Large {\bf Highly Damped Quasinormal Modes and the Small Scale Structure of Quantum Corrected Black Hole Exteriors}}
\vspace{1cm} 

%\vfill

\renewcommand{\baselinestretch}{1}
{\bf
James Babb${}^\dagger$,
Ramin Daghigh${}^\sharp$
and
Gabor Kunstatter${}^\dagger$, 
\\}
\vspace*{0.7cm}
{\sl
${}^\dagger$ 
Department of Physics, 
The University of Winnipeg,\\
515 Portage Avenue, 
Winnipeg, Manitoba, Canada R3B 2E9\\
{[e-mail: g.kunstatter@uwinnipeg.ca]}\\[5pt]
}
{\sl
${}^\sharp$ 
Natural Sciences Department,
Metropolitan State University, 700 East Seventh Street, Saint Paul, Minnesota, USA 55106\\
{[e-mail: ramin.daghigh@metrostate.edu]}\\ [5pt]
}
\vspace{2ex}
{\bf Abstract}
\end{center}
Quasinormal modes provide valuable information about the structure of spacetime outside a black hole. There is also a conjectured relationship between the highly damped quasinormal modes and the semi-classical spectrum of the horizon area/entropy. In this paper, we show that for spacetimes characterized by more than one scale, the ``infinitely damped" modes in principle probe the structure of spacetime outside the horizon at the shortest length scales. We demonstrate this with the calculation of the highly damped quasinormal modes of the non-singular, single horizon, quantum corrected black hole derived in \cite{pk09}. 
%This calculation illustrates the deep connection between the short scale structure of black hole spacetimes and the highly damped quasinormal modes. It also provides an example of a black hole spacetime for which the analytic technique based on the complex coordinate WKB method\cite{andersson} appears problematic while the slightly more general monodromy technique\cite{motl2} is tractable.
%\vfill \hfill 
%August 2010
\end{titlepage}

\section{Introduction}

During the last decade, a great deal of attention has been given to the ``ringing" or quasinormal mode (QNM) spectra of black holes. QNM's are the damped fundamental frequencies at which black hole spacetimes resonate when perturbed. The long-lived QNM's are in principle observable in the gravitational waves emitted during the ring down of a black hole formation. The highly damped modes, while not in practice observable, have in recent years become a subject of interest due in part to a conjecture by Hod\cite{hod98} relating them to the spacing between semi-classical area/entropy eigenvalues of the quantum black hole. Hod's argument was based upon earlier numerical results for the highly damped QNM's of the Schwarzschild black hole and the Bohr Correspondence Principle. 

Our purpose here is not to settle the controversy over this intriguing conjecture. 
Whether or not one subscribes to a specific relationship between highly damped QNM's and the black hole area spectrum, the impressive body of calculations in the literature makes it clear that there is a deep connection between the black hole QNM's and the small scale structure of the black hole spacetime. The nature of this connection was at first considered a bit mysterious due to the fact that the calculation of the asymptotic QNM's required analytic continuation of the exterior solution to the entire complex $r$-plane. This generated questions along the lines of: ``Why would the vibrational modes exterior to the horizon care about the singularity at the origin?"  An answer to this was already implied in the paper of Andersson and Howls\cite{andersson} in the context of the highly damped QNM's of Reissner-Nordstr$\ddot{\rm o}$m (R-N) black holes. It had been observed that in the highly damped limit, the real part of the QNM's of the R-N black hole approached $\ln(5)$ when the charge $q$ was taken to zero. This appeared to be in contradiction to the fact that for the Schwarzschild solution (which of course is the $q\to 0$ limit of the R-N solution) one obtains $\ln(3)$. Andersson and Howls suggested that the explanation for this apparent contradiction lay in the existence of two separate scales in the R-N problem, both in addition to the scale set by the ADM mass of the solution. They argued that care had to be taken as to the order in which the various scales were taken to zero or infinity. Specifically, the $q\to 0$ limit does not commute with the $|\omega_i|\to \infty$ limit, where $\omega_i$ is the imaginary part of the complex QNM frequency $\omega$ which is also the damping rate. If the former limit is taken first, one obtains $\ln(3)$ whereas if the latter limit is taken first, $\ln(5)$ is the result. They noted therefore that while $\ln(5)$ is the correct R-N result for very high damping, there should be an intermediate range of damping for which the Schwarzschild value of the real frequency is correct. Order of magnitude arguments suggest this range in four spacetime dimensions ($4-d$) to be:
\beq
1<<  GM |\omega|  << (GM)^4/q^4~,
\label{eq:Schwarz damping range}
\eeq
where $M$ is the black hole mass and $G$ is Newton's gravitational constant.  Note that $|\omega| \approx |\omega_i|$ in the large damping limit.  This observation was verified explicitly in \cite{rn_qnm}.  A similar situation also appears in Kerr black holes with small angular momentum. In the case of Kerr spacetime, it was observed by Keshet and Hod\cite{Keshet-Hod} that in the highly damped limit, the real part of the QNM frequency approached zero when the angular momentum per unit mass $a$ was taken to zero. This is again in contradiction to the fact that for the Schwarzschild solution, which is the $a\to 0$ limit of the Kerr solution, one obtains $\ln(3)$.  This issue was resolved by Green, Mulligan and one of us in \cite{DGM-Kerr}, where it was shown that in the intermediate range of
\beq
1<< GM |\omega|  << \frac{GM}{a}~
\label{Schwarz damping range for Kerr}
\eeq
the Schwarzschild value of the real part of the QNM frequency is recovered.
These observations have important implications for interpreting the highly damped QNM's.  

Of course the QNM's cannot, by definition, probe the black hole interiors directly.  However, for each independent scale associated with the black hole exterior, there is an appropriate range of dampings over which the QNM spectrum is sensitive to the structure of the exterior metric at that scale.
It is perhaps worth elaborating this point a bit further. The equations governing the propagation of black hole QNM's can invariably be reduced to a scalar equation of Schr$\ddot{\rm o}$dinger form:
\begin{equation}
\frac{d^2\psi}{dr^2}+R(r)\psi =0~,
\end{equation}
where the ``potential" $R(r)$ depends explicitly on the form of the metric, and on the QNM frequency $\omega$. For example, for pure electromagnetic and gravitational perturbations in the background of a $4-d$ R-N black hole, the potential can be approximated by
\begin{eqnarray}
R(r)   &\approx & \frac{1}{f^2}  \left \{ \omega^2  -\frac {l(l+1)}{r^2}+\frac{j^2+l(l+1)}{r^3} 2GM \right.  \nonumber \\
  &-& \left.  \frac{j^2-1/4}{r^4}(2GM)^2 
       +\frac{j^2+6}{r^5}2GMq^2-\frac{6}{r^6}q^4 \right\}~
\label{R-RN-EM-Grav}
\end{eqnarray}
when $q << GM$.  Here $j$ is the spin of the perturbation, which is $0$ for scalar, $1$ for electromagnetic and $2$ for gravitational perturbations.  In the case of infinite damping, i.e.
\begin{equation}
GM |\omega|>> {\frac{(GM)^4}{q^4}}>>1~,
\end{equation}
the analytic techniques are only consistent if one retains the last term in $R(r)$, despite the fact that exterior to the horizon, $r>2GM$, this term is much less than those that depend only on $M$. That is
\begin{equation}
\frac{q^4/r^6}{(GM)^2/r^4}<\frac{q^4}{(GM)^4}<<1~.
\end{equation}
This is what is meant by the statement that for suitably large damping, the corresponding QNM spectrum is sensitive to the small scale structure of the spacetime. On the other hand, for intermediate damping, as in  (\ref{eq:Schwarz damping range}), one can safely ignore all terms containing $q$ so that the spectrum is purely Schwarzschild and contains no information about corrections due to the presence of electric charge and the inner horizon.  Similar arguments can be made for the Kerr case as well.

%%%%%%%%%%%%%%%%%%%%%%%
The above discussion leads one to ask about the role of the classical singularity in determining the highly damped QNM's. At some scale, presumably the Planck scale, quantum gravity corrections will become important. A great deal of work has recently been devoted to the question of how, or indeed if, the singularity at the center of black holes gets resolved by quantum gravitational effects. Most of this work has been in the context of loop quantum gravity\cite{ashtekar05}-\cite{klp09}. One can therefore ask how the resulting quantum gravity motivated corrections to black hole spacetimes exterior to the horizon affect the highly damped QNM's. Or to put it differently, can one {\bf even in principle} obtain information about the quantum gravity induced small scale structure of a quantum corrected (QC) black hole by measuring its highly damped QNM's? 

The purpose of this paper is to examine this question in the context of the particular QC black hole spacetime derived by Peltola and one of us in \cite{pk09} using an effective polymer quantized theory of the homogeneous Schwarzschild interior. This QC black hole has several properties that make it particularly well suited for such an investigation. First of all, it is possible to explicitly write down a relatively simple analytic expression for the metric. This is a necessary condition for doing the analytic calculations of the highly damped QNM's. Secondly, the solution is completely regular everywhere, but has only a single horizon. The exterior is asymptotically flat and at large distances takes the usual Schwarzschild form.   Using the language of the maximally extended Kruskal spacetime to describe the interior, one finds that the throat of the Einstein-Rosen wormhole contracts to a minimum ``bounce radius" before expanding to infinity. Thus the interior asymptotes to a Kantowski-Sachs type homogeneous but anisotropic spacetime. The singularity is avoided because, in effect, $r=0$ is not in the manifold. It is therefore intriguing to see how the non-singular nature of the interior affects the highly damped QNM's as (in principle) seen in the exterior. 

It should be noted that there are two commonly used methods for the analytic (as opposed to numerical) calculation of asymptotic QNM's. The Motl-Neitzke (MN)\cite{motl2} method was first used to obtain analytic expressions for the highly damped QNM's of  Schwarzschild black holes in four dimensions and higher. These results were later confirmed using a slightly different method by Andersson-Howls (AH)\cite{andersson}. Both methods have since been applied to a large variety of black hole spacetimes
%\cite{qnm_refs}
. In some cases, such as the Gauss-Bonnet black hole, it turns out that the analytic structure of the solution makes it difficult to implement the AH procedure \cite{dkz06}. This is related to the presence of a branch point singularity in the Schwarzschild form of the metric, which appears in the complex plane. The QC black hole that is examined here, despite being non-singular, also has such a branch point singularity in the Schwarzschild form of the metric and a preliminary examination points to potential difficulties with the AH method in this case as well. 
We will show that the MN method can nonetheless be implemented in a rather straightforward fashion. 

The structure of the paper is as follows. In section 2 we review briefly the AH and MN methods for the Schwarzschild black hole, introducing the notion of Stokes and anti-Stokes lines for the uninitiated.   
%We describe in somewhat more detail a variation of the MN method that clarifies how the choice of scale for the damping determines the scale of the ``fine structure" in the solution that the QNM spectrum reveals. 
Section 3 is devoted to a review of the QC black hole of \cite{pk09}, including a presentation of the Stoke/anti-Stokes lines for this case as well as a demonstration of why the AH method is problematic. In section 3, we also present our calculation of the asymptotic QNM's for the QC black hole using the MN method. Finally, we close in section 4 with conclusions and prospects for future work.

%%%%%%%%%%%%%%%%%%%%%%%%%%%%%%%%%%%%%%%%%%%%%%%%%%%%%%%%%%%%%%%%%%%%%%%%%%%%%

\section{Highly Damped Quasinormal Mode Calculations}
\subsection{General Method}
In order to illustrate the method in its simplest form we consider the QNM's of a massless scalar field propagating in a Schwarzschild black hole background. The equations for the gravitational wave modes, after the appropriate tensor decompositions\cite{Ishibashi1}, are qualitatively similar.
The Klein-Gordon equation we wish to consider has the form
\begin{equation}
\partial_\mu{\sqrt{-g}}g_{{\mu}{\nu}}\partial_\nu{\Psi}=0~,
\label{general-wave-eq}
\end{equation}
where $g_{\mu\nu}$ is the metric and $g$ is it's determinant. 
%%%%%%%%%%%%%%%%%%%%%%%%%%%%%%%%%%
In a completely general, spherically symmetric, static spacetime, the line element can be written as
\begin{equation}
ds^2=-f(r)dt^2+g(r)^{-1}dr^2+r^2d\Omega^2~.
\end{equation} 

We separate variables $(r,t)$ as follows
\begin{equation}
\Psi(r,t) = e^{-i\omega t} \frac{\psi(r)}{r \sqrt{F}}~,
\end{equation}
where $F=\sqrt{fg}$ and $\omega=\omega_r + i \omega_i$ is in general a complex frequency such that $\omega_r>0$ and $\omega_i<0$.
The spatial part of the wave function $\psi(r)$ obeys
\begin{equation}
\frac{d^2\psi}{dr^2}+R(r)\psi=0~,
\label{eq:de psi}
\end{equation}
where the effective potential $R(r)$ is
\begin{equation}
R(r)=\frac{\omega^2}{F^2}-\frac{1}{2}{\frac{\partial_r^2F}{F}}+\frac{(\partial_rF)^2}{4F^2}-\frac{2}{r}\frac{\partial_r F}{F}~.
\label{eq:def R}
\end{equation}
 Note that for non-extremal horizons $F$ has a simple zero at the horizon location. Moreover, for asymptotically flat spacetimes, $F\to 1$ as $r\to \infty$ so the potential $R\to \omega^2$. The asymptotic solutions are therefore ingoing and outgoing spherical plane waves, as expected.

The QNM's are defined as usual to be solutions to the wave equation that are purely ingoing at the (outer) horizon and outgoing at spatial infinity. In the limit that the magnitude of the imaginary part of $\omega$ is large (this will be made more precise in what follows), the only terms in $R(r)$ other than the term containing $\omega^2$, which are relevant to the calculation, are those that dominate near any poles that may be present in $R$. 

In the case of the Schwarzschild solution, the line element is given by
\begin{equation}
ds^2=-(1-\frac{2GM}{r})dt^2+(1-\frac{2GM}{r})^{-1}dr^2+r^2d\Omega^2~.
\end{equation}
In this case there are poles in $R(r)$ only near the horizon and the origin. In the infinite damping limit, the relevant part of the  potential is simply
\begin{equation}
R(r)\approx \frac{\omega^2}{F^2}-\frac{4j^2-1}{4r^2}~,
\label{simpleR}
\end{equation}
where $j=0$ for scalar perturbations.
%%%%%%%%%%%%%%%%%%%%%%%%%%%%%%%%%%%%%%%%%%%%%%%%%%%%%

The analytic calculation is based on the  Wentzel-Kramers-Brillouin (WKB) approximation, which gives the two  linearly-independent solutions to be
\begin{equation}
f_{1,2}^{(t)}(r)=Q^{-\frac{1}{2}}(r)\exp\left[\pm{i}\int ^{r}_{t}Q(r')dr'\right]~,
\label{WKB-solutions}
\end{equation}
where $t$ is one of the zeros of the function $Q(r)$.
%, which is introduced in order to guarantee that the WKB solution is accurate near the poles of $R$\cite{andersson}.
Generally, one needs to take 
\begin{equation}
Q^2(r)=R(r)-\sum_{P}\frac{1}{4(r-P)^2}~,
\label{QvsR}
\end{equation} 
where $P$ are the location of the poles of $R$. We define the square root of $Q^2$ in the WKB solution so that the positive exponential term (i.e. $f_1$) is outgoing at infinity. 

%We have one pole at the origin (as well as one at the event horizon which we do not need to worry about for the moment). This, in the present case, near the pole at the origin we end up with:
%\begin{equation}
%Q^2(r)={\frac{r^2}{4M^2}}[\omega^2-\frac{15M^2}{r^4}]-\frac{1}{4r^2}
%\end{equation}

%Note that for large $r$, $Q \to const$ so that given the time dependent ansatz \ref{time dependent}, if we assume that the real part of the frequency is positive, $f_1$ is the outgoing solution asymptotically.

\begin{figure}[th]
	\centering
		 \includegraphics[height=9.cm]{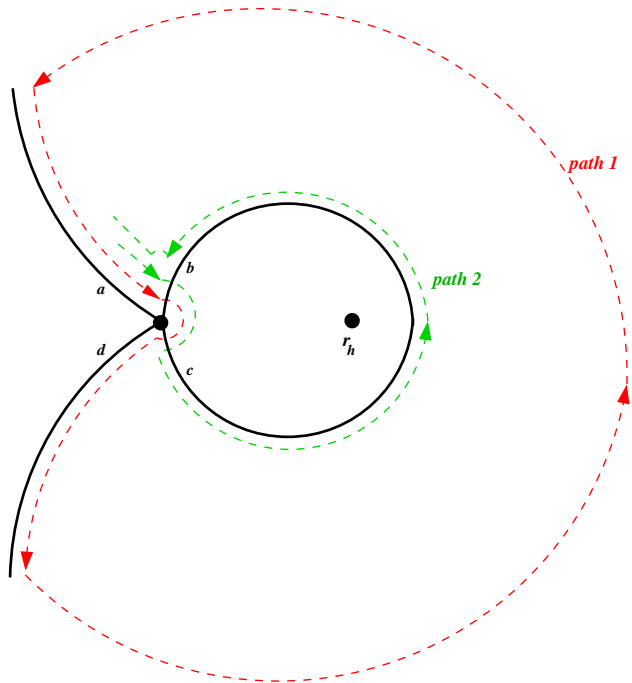}
	\caption{\footnotesize A simplified schematic illustration of anti-Stokes lines far from the origin for the Schwarzschild spacetime in the complex $r$-plane.  The filled circles are the poles of the function $Q(r)$ at the event horizon ($r=r_h$) and the origin ($r=0$).  The dashed lines represent the paths that we follow along anti-Stokes lines to determine the QNM spectrum.}
\label{fig:far Schwarzschild}
\end{figure}

The basic idea of both the AH and MN analytic methods is to find a closed contour around the pole at the event horizon in the complex $r-$plane along which the WKB approximation remains valid and evaluate the change in the phase of the above expressions (the monodromy) along this contour. If the contour can be suitably connected to $r=\infty$, one can impose the boundary condition that the solution be purely outgoing ($f_1$) there as a ``starting point" and see how the solution changes as one goes around the contour. If in addition the chosen contour can be deformed without obstruction to a contour close to and encircling the event horizon, one can impose the second boundary condition that the solution is purely ingoing at the horizon to evaluate the monodromy along this contour. Requiring the two monodromies to be equal gives the WKB condition on the QNM frequency and determines the spectrum. A useful concept in choosing the contours is the anti-Stokes line, which is a line along which $Q(r)dr$ is purely real and thus our WKB solutions are periodic functions.  The anti-Stokes lines far from the origin ($r=0$) and two suitable contours ({\it path 1} and {\it path 2}) for our calculations are shown in Fig.\ \ref{fig:far Schwarzschild}.  Note that {\it path 2} starts at point $a$.  Then it moves to point $c$ and after encircling the horizon it returns back to point $a$.  This was the loop chosen by Andersson and Howls{\footnote{Note that Andersson and Howls encircle the horizon in the clockwise direction in \cite{andersson}, but {\it path 2} runs in the counter-clockwise direction.  This difference should not and does not affect the results.}}.  The other contour, {\it path 1}, was the one used by Motl and Neitzke. It starts at point $a$ in Fig.\ \ref{fig:far Schwarzschild}.  Then, it moves to point $d$ and follows the line shown in the figure to infinity and circles around at infinity in the counter-clockwise direction to meet the line in the upper half plane and finally it returns back to point $a${\footnote{Note that, in \cite{motl2}, Motl and Neitzke start their loop at point $d$ and rotate in the clockwise direction along {\it path 1}.  The reason for this difference is the fact that Motl and Neitzke assume that perturbations depend on time as $e^{i\omega t}$, which means $\omega_i>0$.  In this paper we assume that perturbations depend on time as $e^{-i\omega t}$, which means $\omega_i<0$.}}. Since {\it path 1} can be contracted to {\it path 2} without encountering any poles, one expects them to yield equivalent results, which indeed they do. %The infinite contour is the one chosen by MN but their method could also have been used on the finite contour ($path\, 2$). The answers must agree since the infinite and finite contours differ by a contour that does not enclose any poles and therefore has trivial monodromy. 

%The disadvantage of {\it path 1} is that, as pointed out by Motl and Neitzke, it can only be trusted near infinity for the dominant mode, so that the subdominant mode must be dropped. While this is sufficient to give the correct condition on the frequency, one gives up the ability to do a consistency check using the monodromy of the subdominant mode, as we did with the AH contour in the previous section.  

 As long as one stays on an anti-Stokes line the relative weights of the two WKB solutions do not change. So if the solution is purely $f_1$, it remains $f_1$ as long as one does not leave the anti-Stokes line. The problem, as illustrated by the two contours in Fig.\ \ref{fig:far Schwarzschild} is that one cannot find a suitable closed contour that remains on a single anti-Stokes line, since in general they emanate from the neighborhood of poles in $R(r)$. The trick is to know how the solution changes as one ``jumps" from one anti-Stokes line to another in the neighborhood of a pole.  
%There are basically two methods in the literature for doing this. The first general calculation of the asymptotic QNM's was done by 
Motl and Neitzke\cite{motl2} solved the approximate wave action analytically
near the poles in order to calculate the monodromy along the contours in their vicinity.  Andersson and Howls\cite{andersson} clarified the MN calculation by using the so-called Stokes phenomena to calculate the monodromies in the vicinity of the poles. In all previous calculations in the literature, both methods gave identical results, as they must. 

We will now describe both in the context of the Schwarzschild black hole before proceeding to apply them to the non-singular black hole of reference \cite{pk09}.

\subsection{The Andersson-Howls Method Applied to Schwarzschild Black Holes}
\begin{figure}[th]
\centering
\includegraphics[height=9.cm]{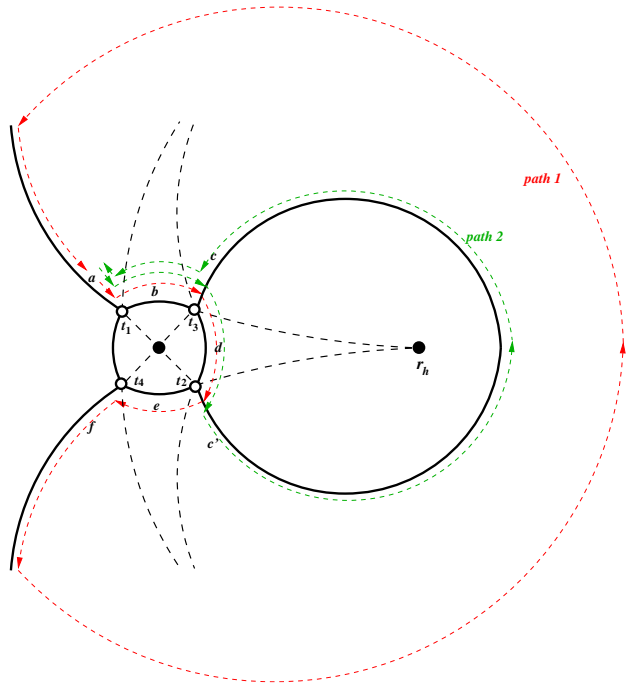}
\caption{\footnotesize A complete schematic illustration of Stokes (dashed) and anti-Stokes (solid) lines for the Schwarzschild spacetime in the complex $r$-plane.  The hollow circles are the zeros and the filled circles are the poles of the function $Q(r)$.  The poles are located at the event horizon ($r=r_h$) and the origin ($r=0$).  The paths ({\it path 1} and {\it path 2}) that we can follow along anti-Stokes lines to determine the QNM spectrum are represented by thin dashed lines with lighter colors.}
\label{fig:Schwarzschild}
\end{figure}

%One of the analytical methods used to evaluate a black hole's quasinormal mode spectrum was developed by Andersson and Howls in 2004[4].  
The AH method requires a detailed knowledge of the analytic structure of the function $Q(r)$ near the pole at $r=0$. 
For this, one needs to determine the location of the zeros of $Q$ in the complex $r$-plane. It is important to do this as it is from the zeros that the Stokes and anti-Stokes lines emanate. There are three of each emerging from each zero in the complex plane, being initially spaced $2\pi/3$ radians apart from one another.  We also need to introduce a branch cut from each zero to ensure that $Q$ remains single-valued.  These cuts need to be introduced in a way that they do not affect our analysis.  Specifically, they should not intersect {\it path 1} and {\it path 2} in Fig.\ \ref{fig:Schwarzschild}. The Stokes lines are the lines in the complex plane where $Q(r)dr$ is purely imaginary (and thus one of the WKB solutions $f_{1,2}^{(t)}$ blows up exponentially as one moves away from the zero of the function $Q(r)$ at $t$ along Stokes lines).
%, and the anti-Stokes lines are where $Q(r)dr$ is purely real (and thus our WKB solutions are periodic functions)
When one moves from one anti-Stokes line to an adjacent anti-Stokes line in the vicinity of a zero, a Stokes line has to be crossed.  In this case, Stokes phenomenon dictates that the WKB solution, which is dominant on the Stokes line, be replaced by the dominant WKB solution $\pm i$ times the sub dominant term.  The positive sign is used when one crosses the Stokes line in the anti-clockwise direction and the negative is used for the clockwise crossing.   Note that such simple change, i.e. the addition of $\pm i$ times the sub dominant WKB solution to the dominant one, only holds when the Stokes line emanates from the zero, which is used as the lower limit in the phase-integral that appears in the WKB solutions $f_{1,2}^{(t)}$.  This means that in order to construct an approximate solution valid in different regions of the complex plane we need to switch the lower limit of our phase integral between zeros.  This requires the evaluation of integrals of the type
\begin{equation}
\gamma_{ij}=\int_{t_i}^{t_j}Q(r)dr~,
\label{gamma-ij}
\end{equation}
where $t_i$ and $t_j$ are two zeros of $Q$.  
%%%%%%%%%%%%%%%%%%
%%%%%%%%%%%%%%%%%%%
To derive a WKB condition on QNM's, we can take either {\it path 1} or {\it path 2}.  In both paths, we start on line labeled $a$ on which we know what the solution is due to the boundary condition at infinity:
\begin{equation}
\Psi_a=f_1^{(t_1)} ~,
\end{equation}
where the superscript $t_1$ indicates that the WKB phase integral is to be evaluated from the zero $t_1$ shown in Fig.\ \ref{fig:Schwarzschild}.  In the case of {\it path 1}, we can move along anti-Stokes lines, while applying Stokes phenomenon when we jump from one anti-Stokes line to the other near zeros, to the line labeled $f$ that extends to infinity in the lower half plane.  On line $f$, we get the solution
\begin{equation}
\Psi_{f}= -(e^{-3i\gamma}+e^{-i\gamma}+e^{i\gamma})f_1^{(t_4)}+i(e^{-3i\gamma}+e^{-i\gamma}+e^{i\gamma}+e^{3i\gamma})f_2^{(t_1)}~,
\end{equation}  
where
\begin{equation} 
\gamma=\gamma_{31}=\gamma_{32}=\gamma_{42}~,
\end{equation} 
which can be evaluated analytically in the limit $r\rightarrow 0$.  In this limit, we can use Eqs.\ (\ref{simpleR}), (\ref{QvsR}), and (\ref{gamma-ij}) to show that 
\begin{equation}
\gamma_{31}\approx \int_{t_3}^{t_1}\left[{\frac{r^2\omega^2}{(-2GM)^2}}-\frac{j^2}{r^4}\right]^{\frac{1}{2}}dr={\frac{j}{2}} \pi=\gamma~.
\end{equation}

We can now close {\it path 1} at infinity, where $f_1$ is dominant.  %($f_2$ is exponentially small in this part of the contour, and therefore its coefficient should not be trusted.)  
After returning back to line labeled $a$, we get
\begin{equation}
\Psi_{\bar{a}}= -(e^{-3i\gamma}+e^{-i\gamma}+e^{i\gamma})e^{i(\Gamma_{ccw}+\gamma)}f_1^{(t_1)}+\mbox{sub dominant term}~,
\end{equation}  
where $ccw$ stands for counter-clockwise and
\begin{equation}
\Gamma_{ccw}=\mathop{\oint}_{ccw}{Q}dr=2\pi{i}\mathop{Res}_{r=2M}(Q)=4\pi i\omega GM.
\end{equation}
We now can use the fact that
\begin{equation}
\Psi_{\bar{a}}= e^{-i\Gamma_{ccw}}\Psi_a~,
\label{Schwarzschild-monodromy}
\end{equation}
where $e^{-i\Gamma_{ccw}}$ is the monodromy of our loop according to the boundary condition at the event horizon to get the WKB condition
\begin{equation}
e^{-2i\Gamma_{ccw}}= e^{8\pi \omega GM}=-1-2\cos(2\gamma)=-1-2\cos(j\pi)~
\label{Schwarzschild-WKB}
\end{equation} 
on the QNM frequency $\omega$.  The disadvantage of {\it path 1} is that, as pointed out by Motl and Neitzke, it can only be trusted near infinity for the dominant mode, so that the sub dominant mode must be dropped. While this is sufficient to give the correct condition on the frequency, one gives up the ability to do a consistency check using the monodromy of the sub dominant mode.  The AH contour ({\it path 2}), however, does not suffer from such disadvantage.  
In the case of {\it path 2}, after circling around the horizon, we get back to the line labeled $a$, where we get
\begin{equation}
\Psi_{\bar{a}}= -(1+e^{2i\gamma}+e^{-2i\gamma})e^{i\Gamma}f_1^{(t_1)}-i(1+e^{2i\gamma})\left[e^{-i\Gamma}+e^{i\Gamma}(1+e^{2i\gamma}+e^{-2i\gamma})\right]f_2^{(t_1)}~.
\end{equation}  
Comparing $\Psi_{\bar{a}}$ and $\Psi_a$ tells us that the coefficient of the sub dominant WKB solution $f_2$ has to be zero.  This gives us the same WKB condition as in Eq.\ (\ref{Schwarzschild-WKB}).  We now can use (\ref{Schwarzschild-WKB}) to show that the coefficient of the dominant WKB solution $f_1$ is equal to the correct monodromy presented in Eq.\ (\ref{Schwarzschild-monodromy}).  This is what we mean by consistency check.

\subsection{The Motl-Neitzke Method Applied to Schwarzschild Black Holes}
An alternative means of determining the QNM spectrum is given by Motl and Neitzke. For this method the simplified version of the anti-Stokes line topology in the infinite damping limit, shown in Fig.\ \ref{fig:far Schwarzschild}, is sufficient.  In other words we can ignore the ``fine structure" of zeros surrounding the pole at the origin, as well as the details of the Stokes lines.
% we may do this (as we will see) because we need never get close enough to the pole for these zeros to be an issue.  
The basic idea is to approach each singularity in the complex plane along an anti-Stokes line closely enough so that a suitable approximation to the differential equation (\ref{eq:de psi}) can be solved analytically. In all cases that have been considered so far, the solution near the poles is well approximated by a Bessel function. This approximate analytic solution is then used instead of the Stokes phenomenon to ``jump" between anti-Stokes lines near a pole.  

Our task then is, essentially, to ``walk" our solution around a closed contour such as {\it path 1} or {\it path 2} using the WKB approximation along the anti-Stokes lines and the approximate analytic (Bessel function) solution near the poles.

We now calculate the monodromy around {\it path 1} using the MN method. Note that we have chosen the branch cut from the origin to point along the negative real axis. We start at the point $a$ in Fig. \ref{fig:far Schwarzschild}, such that $r_a<<2GM$ which is near enough to the pole at $r=0$ so that Eq.\ (\ref{eq:de psi}) can be well approximated by  
\begin{equation}
\frac{d^2\Psi}{dr^2}+\left[{\frac{r^2\omega^2}{(-2GM)^2}}-\frac{4j^2-1}{4r^4}\right]\Psi=0~.
\end{equation}
The general solution to the above is:
\begin{equation}
\Psi=C_{1}\sqrt{\pi r}J_{\frac{j}{2}}(z)+C_{2}\sqrt{\pi r}J_{-{\frac{j}{2}}}(z)~,
\end{equation}
where 
\begin{equation}
z=\frac{r^2\omega}{-4GM}
\label{eq:z def}
\end{equation} and $J_{\pm \nu}$ are Bessel functions of the first kind.  We now need to relate this general solution, near the pole at $r=0$ along the anti-Stokes line labeled $a$ to the outgoing and ingoing WKB wave solutions (\ref{WKB-solutions}).  Sufficiently far from the pole at $r=0$ these are well approximated by $f_1^{(0)}\approx \sqrt{\frac{-2GM}{r\omega}}e^{iz}$ and $f_2^{(0)}\approx \sqrt{\frac{-2GM}{r\omega}} e^{-iz}$ respectively.
Next we evaluate the asymptotic form of the Bessel functions. 
To do this, we use the fact that
\begin{equation}
J_{\pm \nu}(z)=z^{\pm \nu}\phi(z)~,
\label{holomorphic-bessel}
\end{equation}
where $\phi$ is an even holomorphic function.  This means we can write
\begin{equation}
J_{\pm\nu}\left(e^{i \theta} e^{-i \theta}z\right)=e^{\pm i\nu\theta}J_{\pm\nu}\left(e^{-i\theta}z\right)~,
\end{equation}
where $\theta=\arg{z}$.
Note that $e^{-i\theta}z=|z|$ is positive and real.  As long as we stay away from $r=0$ at a distance where
\begin{equation}
|r|>> \sqrt{\frac{4GM}{|\omega|}}~,
\end{equation}
we can assume that $|z|>>1$.  This allows us to use the asymptotic behavior of the Bessel function
\begin{equation}
J_{\pm\nu}(x)\rightarrow\sqrt{\frac{2}{\pi{x}}}\cos\left(x\mp \nu \frac{\pi}{2}-\frac{\pi}{4}\right)~\mbox{for}~x>>1
\label{asymptotic-bessel}
\end{equation}
to show that 
\begin{eqnarray}
\sqrt{\pi r}J_{\pm\nu}(z) &=& \sqrt{\pi r}e^{\pm i \nu\theta}J_{\pm \nu}(e^{-i\theta}z) \nonumber  \\
& \approx & 2  \sqrt{\frac{-2GM}{r\omega}} e^{i {\frac{2\theta}{\pi}}\alpha_\pm} \cos(e^{-i\theta} z-\alpha_\pm)~,
\label{asymptotic-bessel-1}
\end{eqnarray}
where $\alpha_\pm\equiv \frac{\pi}{4}\pm\nu\frac{\pi}{2}$. 
If we choose the branch $-\pi<\theta\le\pi$, the argument of $z$ along the anti-Stokes line labeled $a$ is $\theta=2\pi$. 
After substituting $2\pi$ for $\theta$ in the above equation, we find the wave function along anti-Stokes line $a$ in terms of the ingoing and outgoing WKB modes, which is
\begin{equation}
\Psi_a=\left(C_1e^{3i\alpha_+}+C_2e^{3i\alpha_-}\right)f_1^{(0)}(z_a)+\left(C_1e^{5i\alpha_+}+C_2e^{5i\alpha_-}\right)f_2^{(0)}(z_a)~.
\end{equation}
The boundary condition at infinity tells us that, along anti-Stokes lines that extend to infinity, the solution must be dominated by the outgoing mode.  In other words
\begin{equation}
C_1 e^{5i\alpha_+}+C_2e^{5i\alpha_-}=0~.
\label{MN-Sch-condition1}
\end{equation}
We now repeat the above expansion near the point $b$. The anti-Stokes line $b$ is rotated relative to line $a$ one by $-\pi/2$ in the $r$-plane, which means $z_b=e^{-i\pi}z_a$.  Therefore, $\theta$ should be replaced by $\pi$ in Eq.\ (\ref{asymptotic-bessel-1}). This yields, after a few steps:
\begin{eqnarray}
\Psi_b = \left(C_1e^{3i\alpha_+}+C_2e^{3i\alpha_-}\right)f_1^{(0)}(z_b)+\left(C_1e^{i\alpha_+}+C_2e^{i\alpha_-}\right)f_2^{(0)}(z_b)~.
\end{eqnarray} 
Repeating this to get to $c$ and then finally to $d$, we find that ultimately the $-3\pi/2$ rotation in the complex $r$-plane that takes us from the point $a$ on the first anti-Stokes line to the point $d$ on the last anti-Stokes line yields
\begin{equation}
\Psi_d=\left(C_1e^{-i\alpha_+}+C_2e^{-i\alpha_-}\right)f_1^{(0)}(z_d)+\left(C_1e^{-3i\alpha_+}+C_2e^{-3i\alpha_-}\right)f_2^{(0)}(z_d)~.
\end{equation}
The trick now is to realize that we can extend the expression for $\psi_d$ and evaluate it in terms of $z_a$ by using the following integral expressions that are understood to be evaluated along the large loop at infinity:
\begin{equation}
e^{\pm iz_d}= \exp(\pm i\int_0^{r_d}Qdr)= \exp(\pm i \mathop{\oint}_{ccw}{Q}dr) \exp(\pm i\int_0^a Q dr)= e^{\pm i\Gamma_{ccw}}  e^{\pm i z_a}
\end{equation}
so that
\begin{equation}
\Psi_d = \left(C_1e^{-i\alpha_+}+C_2e^{-i\alpha_-}\right)e^{i\Gamma_{ccw}} f_1^{(0)}(z_a)+\left(C_1e^{-3i\alpha_+}+C_2e^{-3i\alpha_-}\right)e^{-i\Gamma_{ccw}}f_2^{(0)}(z_a)\equiv \Psi_{\bar{a}} ~.
\end{equation}
These two expressions $\Psi_{\bar{a}}$ and $\Psi_a$ at $z_a$ must differ by the appropriate monodromy of the WKB phase, which can easily be evaluated by invoking the boundary conditions. In particular, the {\it path 1} can be shrunk without obstruction to a small loop encircling the horizon. Given that the solution near the horizon is ingoing, it must be true that
\begin{equation}
\Psi_{\bar{a}}=e^{-i\Gamma_{ccw}}\Psi_a~.
\label{MN-Sch-WKB-condition}
\end{equation}
The above condition, applied to the coefficient of the dominant solution, yields the following equation:
\begin{equation}
 \left(C_1e^{-i\alpha_+}+C_2e^{-i\alpha_-}\right)e^{i\Gamma_{ccw}}= e^{-i\Gamma_{ccw}} \left(C_1e^{3i\alpha_+}+C_2e^{3i\alpha_-}\right)~.
\label{MN-Sch-condition2}
\end{equation}
Equations (\ref{MN-Sch-condition1}) and (\ref{MN-Sch-condition2}) are two linear equations on the two initial coefficients $C_{1,2}$, which can only be satisfied if the corresponding determinant vanishes. This yields precisely the WKB condition (\ref{Schwarzschild-WKB}) as before.

Note that with the large loop ({\it path 1}) we cannot look at the condition on the sub-dominant mode, so it does not yield any further conditions. However, by looking at the same calculation using {\it path 2}, one can find two different sets of linear equations using the dominant and sub-dominant modes.  Both of these sets of equations in turn should give the same WKB condition in (\ref{Schwarzschild-WKB}).  This provides a very useful consistency check of the calculation, as will be seen in more details when we consider the more complicated case of the QC black hole.

\section{The Quantum Corrected Black Hole}
The QC black hole we wish to consider was derived in \cite{pk09} using a loop quantum gravity inspired quantization scheme similar to those that were applied to black hole interiors by a variety of authors\cite{ashtekar05, modesto06, boehmer07, pullin08}. The basic idea is to start with homogeneous slicings of spherically symmetric spacetimes in Einstein gravity. Polymer quantization\cite{afw,halvorson} is then applied to the resulting Hamiltonian to produce a set of effective, QC equations for the geometrical variables. These equations are solved to find the QC metric in the Schwarzschild interior. In contrast to the previous work, reference \cite{pk09} only applied polymer quantization to the phase space variable corresponding to area. This resulted in a simple analytic expression for the interior metric that could be analytically continued to the exterior in order to construct the complete QC spacetime. The metric for the QC black hole in Schwarzschild-like coordinates is: 
\begin{equation}
ds^2=-\left(\epsilon\sqrt{1-\frac{k^2}{r^2}}-\frac{2GM}{r}\right)dt^2+\frac{dr^2}{\left(\epsilon\sqrt{1-\frac{k^2}{r^2}}-\frac{2GM}{r}\right)\left(1-\frac{k^2}{r^2}\right)}+r^2d\Omega^2~,
\label{qc metric}
\end{equation}
where $k$ is a new (quantum) length scale, and $\epsilon$ can be 1 or -1.

The above metric has intriguing properties. It contains  a single horizon at $r_h=\sqrt{(2GM)^2+k^2}$ so that the analytically continued spatial slices are topologically equivalent to the Einstein-Rosen wormhole. As with Einstein gravity, the radius of the throat of the wormhole is time dependent. The maximum is $r_h$, but instead of shrinking to a singularity in the black hole interior in finite time, the throat radius reaches a minimum at $r=k$, before re-expanding indefinitely to a Kantowski-Sachs spacetime. The resulting complete, single-horizon, non-singular black hole spacetime scenario is therefore reminiscent of past proposals for universe creation via quantum effects inside a black hole\cite{frolov90}.   

To find the QNM's we insert the QC metric (\ref{qc metric}) into Eq.\ (\ref{general-wave-eq}). Using the substitution $\chi=\sqrt{r^2-k^2}$ and extending the domain of $\chi$ to $(-\infty<\chi<\infty)$ (which effectively negates the need for $\epsilon$), we obtain the differential equation
\begin{equation}
\frac{d^2\psi}{d\chi^2}+\tilde{R}(\chi)\psi=0~,
\label{de-qnm-pk}
\end{equation}
where $\psi$ is our rescaled wavefunction and our new potential function $\tilde{R}(\chi)$ is given by
\begin{eqnarray}
\tilde{R}(\chi)&=&\frac{\omega^2(\chi^2+k^2)}{(\chi-2GM)^2}+\frac{k^4-6\chi^2k^2}{4\chi^2(\chi^2+k^2)^2}+\frac{GM}{(\chi^2+k^2)(\chi-2GM)}\nonumber \\
&-&\frac{GM}{\chi^2(\chi-2GM)}-\frac{\chi}{(\chi^2+k^2)(\chi-2GM)}
+\frac{(GM-\chi)^2}{\chi^2(\chi-2GM)^2}~.
\end{eqnarray}
	We immediately notice that this function has poles at $\chi=0$ and $\chi=\pm{ik}$. It also has a pole at the point $\chi=2GM$, which corresponds to the event horizon.  Note that in the infinite damping limit, the relevant part of this potential is simply
\begin{equation}
\tilde{R}(\chi)=\frac{\omega^2(\chi^2+k^2)}{(\chi-2GM)^2}+{\frac{1}{\chi^2}}-{\frac{7k^2}{4(\chi^2+k^2)^2}}~.
\label{R-approx-PK} 
\end{equation}
This is because the last two terms on the right hand side of the above equation dominate all the other terms near the poles whereas the term containing $\omega$ cannot be neglected elsewhere.	 
%Also consider that we are looking at this function in the so-called "high damping limit," where, following Hod's assumption that the imaginary part of ? represents the frequency's damping, we can treat $\omega$ as being an infinitely large, imaginary, negative number. Thus, it is clear that the term $\frac{\omega^2(\chi^2+k^2)}{(2GM-\chi)^2}$ will dominate the function $\tilde{R}(\chi)$ everywhere except in the immediate vicinity of one of the poles (at which one or more of the other terms will no longer be negligible). 
The two linearly-independent WKB solutions to this equation are given by (\ref{WKB-solutions}) with
\begin{equation}
{\tilde Q}^2(\chi)=\tilde{R}(\chi)-\frac{1}{4{(\chi+ik)}^2}-\frac{1}{4\chi^2}-\frac{1}{4{(\chi-ik)}^2}-\frac{1}{4{(\chi-2GM)}^2}~.
\label{Q-PK}
\end{equation}
The last term above can be neglected for large damping rates.
%Note that none of these extra terms will affect the calculation in the vicinity of any pole save the one that it applies to (being insignificant in all other cases). We can then approximate $Q^2(\chi)$ in the vicinity of each of the poles, by throwing out all terms which do not either have $\omega$ in the numerator or a pole in the denominator. Thus reduced, we can easily find the points at which $Q^2(\chi)$ goes to zero in the vicinity of each pole in the complex plane, as well as all Stokes and anti-Stokes lines issuing from each zero. Doing this, we obtain:

Finally, we want to point out that in an intermediate damping region of 
\begin{equation} 
1<< GM|\omega| << {\frac{(GM)^2}{k^2}}~
\label{intermediate-w-range}
\end{equation}
the potential (\ref{R-approx-PK}) reduces to the Schwarzschild potential given in Eq.\ (\ref{simpleR}).  Therefore, in the  above intermediate region of the spectrum, the Schwarzschild value of $\ln(3)$ is expected to appear in the real part of the QNM frequency $\omega$. 

%%%%%%%%%%%%%%%%%%%%%%%%%%%%%%%%%%%%%%%%%%%%%%%%

\begin{figure}
\centering
\includegraphics[height=8.cm]{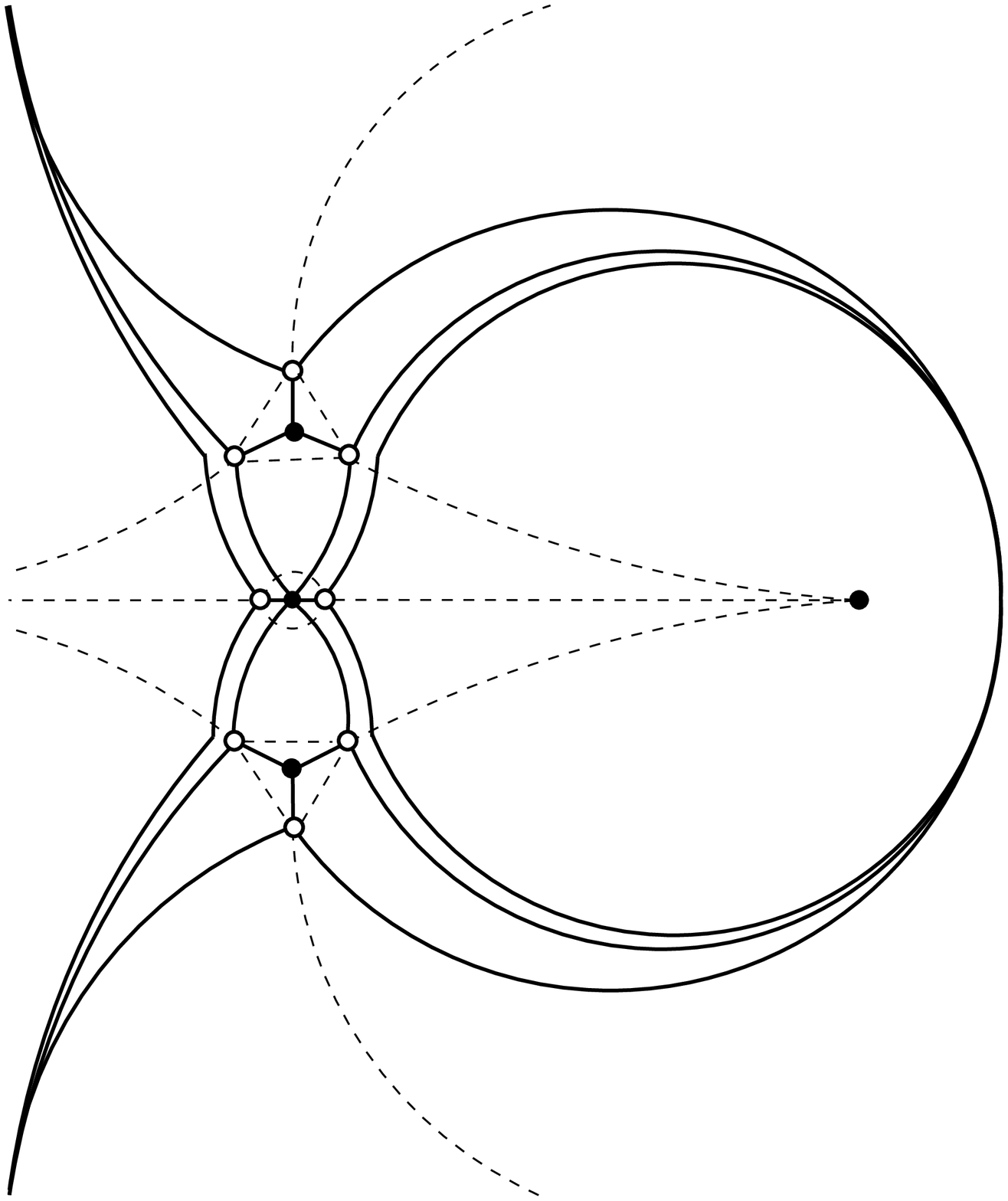}
\includegraphics[height=8.cm]{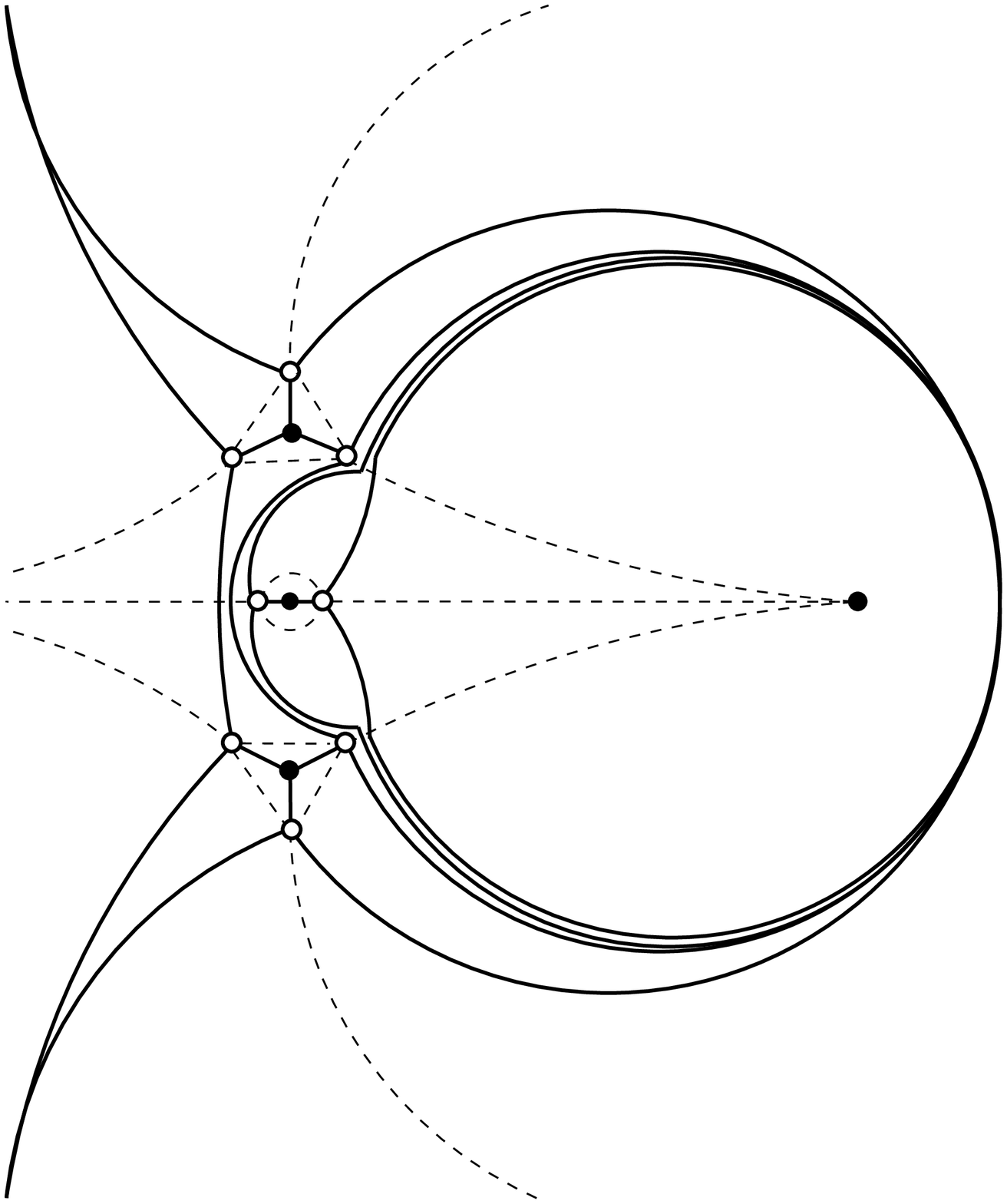}
\caption{\footnotesize A complete schematic illustration of Stokes (dashed) and anti-Stokes (solid) lines for the QC spacetime in the complex $\chi$-plane when $|\omega| \lesssim {\frac{(2GM)^2}{k^3}}$ (left) and $|\omega| > {\frac{(2GM)^2}{k^3}}$ (right).  The hollow circles are the zeros and the filled circles are the poles of the function $Q(\chi)$.}
\label{schem-stokes-lines-ah-pk-large-w}
\end{figure}

\begin{figure}[tb]
\begin{center}
\includegraphics[height=6.5cm]{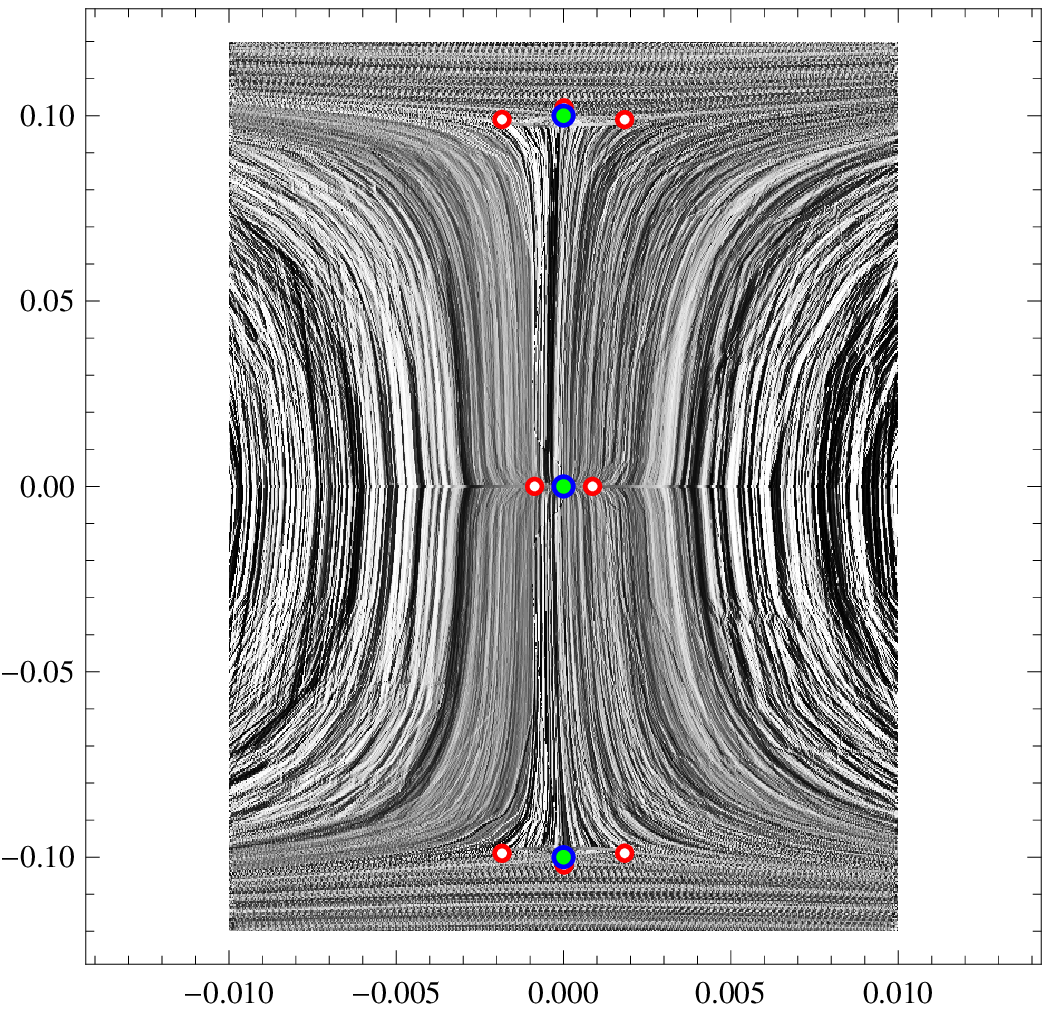}
\includegraphics[height=6.5cm]{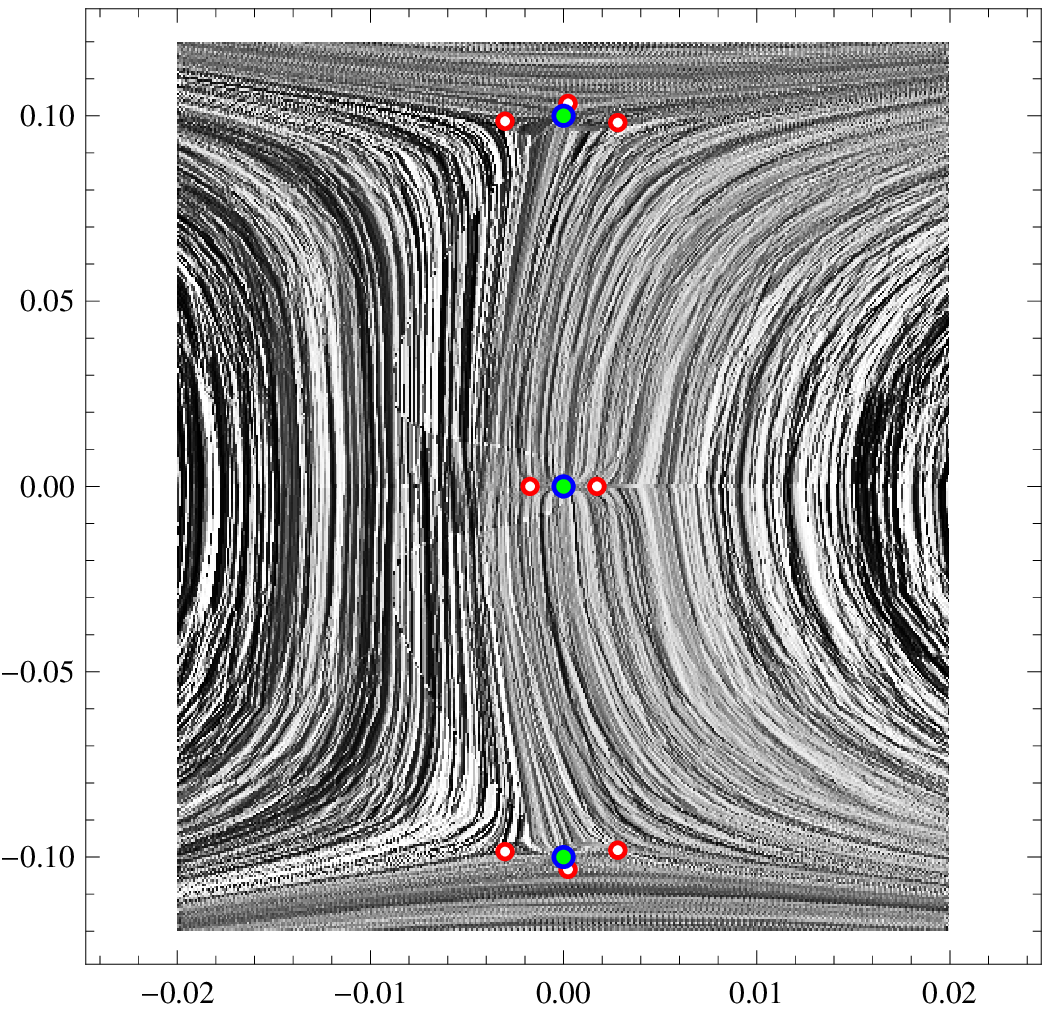}
\end{center}
\caption{\footnotesize Numerically generated anti-Stokes lines for the QC black hole spacetime in $\chi$-coordinate, showing the change in topology as one goes from $|\omega| \lesssim {\frac{(2GM)^2}{k^3}}$ to $|\omega| > {\frac{(2GM)^2}{k^3}}$.  The figure on the left shows the structure of the lines near the origin of the complex plane for $k=0.1$, $2GM=10$ and $\omega=-i 10^5$.   In the figure on the right, $k=0.1$, $2GM=1$ and $\omega=-i 5000$. The filled circles are the poles of $\tilde{Q}$ located at $\pm ik$ and $0$ and the smaller hollow circles are the zeros of $\tilde{Q}$.
%Note that the distance between the zeroes and poles is comparable to the distance between poles at $-ik$, $0$ and $+ik$, so the calculation as done is presumable not very accurate. That's why we have to consider very large $\omega$, for which the topology of the anti-Stokes lines changes. In both cases however we assume that from ``far away" the lines look something like the figure on the left, with vertical Stokes lines extending between the poles, so that our calculation is more or less valid.
}
\label{AH topology change}
\end{figure}

%%%%%%%%%%%%%%%%%%%%%%%%%%%%%%%%%%%%%%%%%%%%%%%%%%%%%%
\begin{figure}
	\centering
		\includegraphics[height=6.5cm]{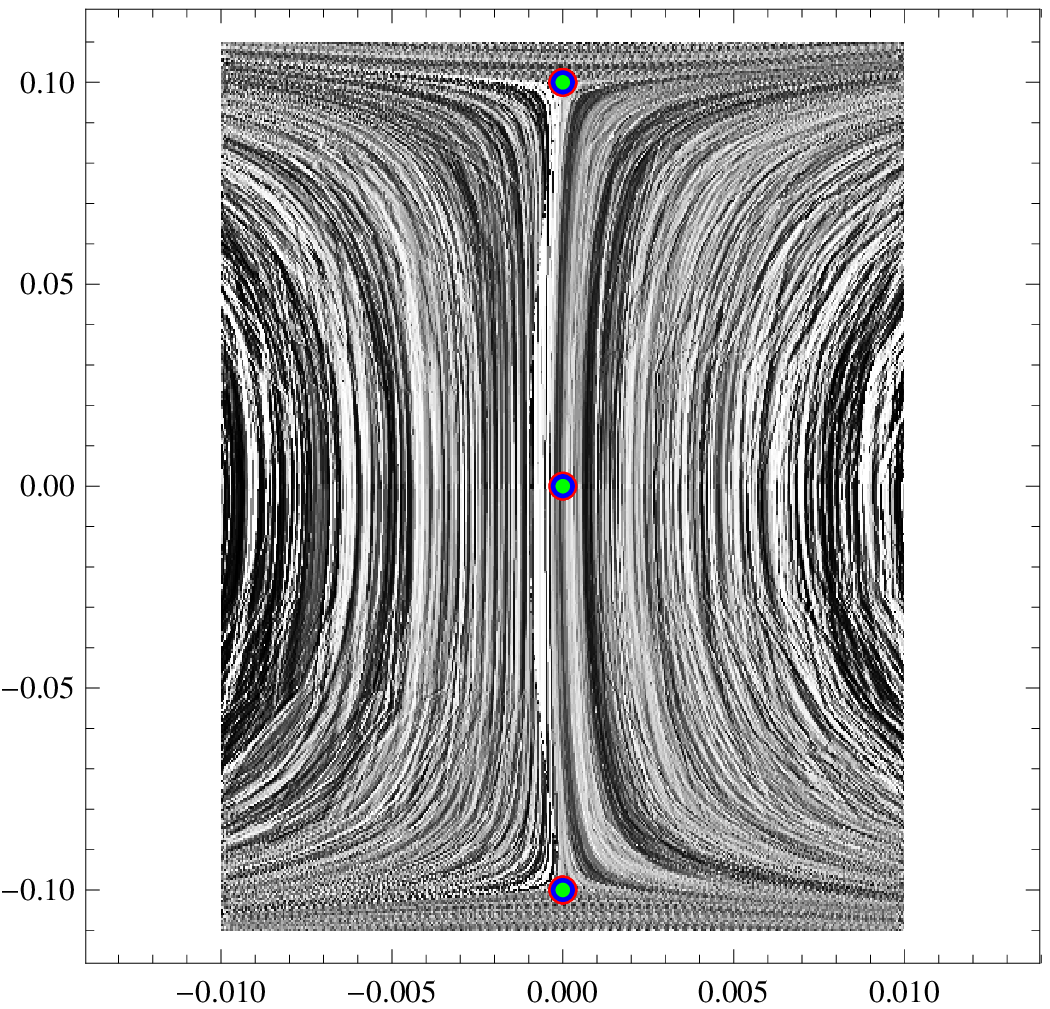}
	\caption{\footnotesize Numerically generated anti-Stokes lines for the QC black hole spacetime in $\chi$-coordinate for $k=0.1$, $2GM=10$ and $\omega=-i10^9$.  The filled circles are the poles located at $\pm ik$ and $0$. 
%The outermost pole represents the black hole's event horizon, with the three poles at left corresponding to those at $-ik$, $0$ and $+ik$.
}
	\label{fig:Stokes3}
\end{figure}

\subsection{The Andersson-Howls Method Applied to the Quantum Corrected Black Hole}
In Fig.\ \ref{schem-stokes-lines-ah-pk-large-w}, we show schematically the topology of Stokes and anti-Stokes lines in two different damping regions for the QC black hole under consideration.  The topology on the right appears in higher damping rates compared to the topology on the left. Fig.\ \ref{schem-stokes-lines-ah-pk-large-w} is plotted based on numerically generated figures of Stokes and anti-Stokes lines, some of which are shown in Figs.\ \ref{AH topology change} and \ref{fig:Stokes3}. After investigating the numerically generated topology of Stokes and anti-Stokes lines for different values of $M$ and $k$, we conclude that in the topology shown on the right in Fig.\ \ref{schem-stokes-lines-ah-pk-large-w} the anti-Stokes line that connects the poles at $\chi=\pm ik$ misses the pole at $\chi=0$ by a minimum distance of $\approx{\frac{k^2}{2GM}}$.  When we lower the damping, the zeros of the function ${\tilde Q}(\chi)$ shown with hollow circles in Fig.\ \ref{schem-stokes-lines-ah-pk-large-w} move away from the poles.  Using Eqs.\ (\ref{R-approx-PK}) and (\ref{Q-PK}), it is easy to show that the distance between the zeros and the pole at $\chi=0$ is $\approx {\frac{2GM}{k|\omega|}}$.  Equating this with the minimum distance that the anti-Stokes line misses the pole at $\chi=0$ leads us to an approximate location of the damping rate where the topology changes from the one on the left to the one on the right in Fig.\ \ref{schem-stokes-lines-ah-pk-large-w}.  This damping rate is 
\begin{equation}
|\omega_i| \approx |\omega| \approx {\frac{(2GM)^2}{k^3}}~. 
\end{equation}

As we have seen during the Schwarzschild case, one ideally needs a closed anti-Stokes contour around the pole at the event horizon, which connects to an outgoing anti-Stokes line.  Critically, however, we require that such a contour never intersects with any pole. A cursory analysis of Fig.\ \ref{schem-stokes-lines-ah-pk-large-w} shows that no such contour exists in either topology. Therefore, this method appears to be problematic for this case.

\subsection{The Motl-Neitzke Method Applied to the Quantum Corrected Black Hole}

In this section, we evaluate the highly damped QNM's of the QC black hole under consideration using the MN method.  In Fig.\ \ref{two-paths}, we show the topology of anti-Stokes lines far from the poles, which is sufficient for applying the MN method. 
As before we construct a simplified contour ({\it path 1} or {\it path 2} in Fig.\ \ref{two-paths}), where our task is to ``walk" our solution from an anti-Stokes line, which extends to infinity, around the pole at the event horizon and back to the same anti-Stokes line  without crossing any of the branch cuts.  As it was pointed out in the last section, 
%the topology of anti-Stokes lines shown in Fig.\ \ref{two-paths} is valid when $|\omega_i| \lesssim {(2GM)^2 \over k^3}$.  
at some damping rate above ${\frac{(2GM)^2}{k^3}}$, the anti-Stokes line that connects the poles at $\chi =\pm ik$ will miss the pole at $\chi=0$.   However, for physical reasons, we are interested in the limit where $k << GM$.  In this limit, the function $\tilde{Q}(\chi)$ can be approximated in the vicinity of the poles at $\chi=0$ and $\chi=\pm ik$, i.e.\ when $|\chi| \approx k$, by
\begin{equation}
\tilde{Q}(\chi) \approx {\frac{\omega \sqrt{\chi^2+k^2}}{-2GM}}~.
\end{equation} 
The above function leads to the topology shown in Fig.\ \ref{two-paths}, where the anti-Stokes line that connects the poles at $\chi =\pm ik$ goes through the pole at $\chi=0$.  This is evident in Fig.\ \ref{fig:Stokes3}, where we have numerically generated  the topology of anti-Stokes lines using the function $\tilde{Q}(\chi)$ in (\ref{Q-PK}), in which the approximate $\tilde{R}(\chi)$ in (\ref{R-approx-PK}) is used, for the case where $k << GM$ ($k=GM/50$).   In the rest of this paper, we do the calculations assuming that $k << GM$.
%In that case our calculations will not be reliable since we cannot form a loop along anti-Stokes lines which excludes all the poles except the pole at the event horizon. 
%Note that the anti-Stokes line that connects the poles at $\chi=\pm ik$ does not go through the pole at $\chi=0$ according to the numerically generated anti-Stokes line topology shown in Fig.\ \ref{fig:Stokes3}.  The minimum distance between this anti-Stokes line and the pole at $\chi=0$ is always less than the ratio $k^2\over 2GM$.  This means that our calculations will not be reliable when ${k^2\over 2GM}\approx 1$ since we cannot form a loop along anti-Stokes lines which excludes all the poles except the pole at the horizon.  However, when $k^2\over 2GM << 1$, it is safe to assume that the anti-Stokes line connecting the poles at $\chi=\pm ik$ goes through the pole at $\chi=0$ as shown in Fig.\ \ref{two-paths}.   
The procedure to solve for the QNM frequencies, for the topology shown in Fig.\ \ref{two-paths}, is as follows:

\begin{figure}[tb]
\begin{center}
\includegraphics[height=9cm]{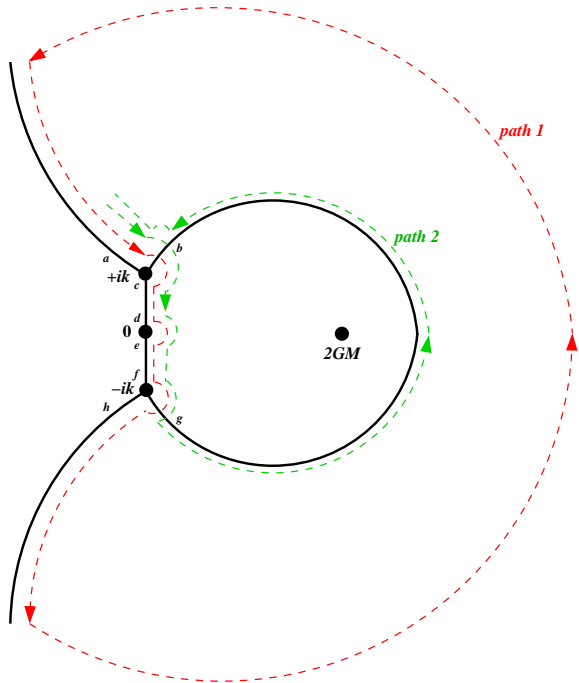}
\end{center}
\caption{\footnotesize A simplified illustration of anti-Stokes lines far from the poles for the QC black hole when $k << GM$.  The outermost pole represents the black hole's event horizon, with the three poles at left corresponding to those at $\pm ik$ and $0$.  The dashed lines are the suitable paths that we follow to determine the QNM spectrum.}
\label{two-paths}
\end{figure}
%From the definition of QNM's, we know that, at the point A, our solution must be purely outgoing. 
We evaluate our solution in the immediate vicinity of the pole at $\chi=ik$.
In this region, differential equation (\ref{de-qnm-pk}) can be expressed as
\begin{equation}
\frac{d^2\psi}{d\chi^2}+\left[\frac{\omega^2(2ik)(\chi-ik)}{(-2GM)^2}+\frac{7}{16(\chi-ik)^2}\right]\psi=0~.
\label{ik-approx-wave-eq}
\end{equation}
This is a Bessel differential equation with a solution of the form
\begin{equation}
\psi=A_+\sqrt{\frac{4\pi}{3}(\chi-ik)}J_{\nu}\left(Z\right)+ A_-\sqrt{\frac{4\pi}{3}(\chi-ik)}J_{-\nu}(Z)~,
\end{equation}
where  $A_{\pm}$ are constants, $\nu=\sqrt{\frac{-1}{12}}$, $J_{\pm{\nu}}$ are Bessel functions of the first kind and
\begin{equation}
Z=\frac{2}{3}\left(\frac{\omega\sqrt{2ik}}{-2GM}\right)(\chi-ik)^{3/2}~.
\end{equation}
$Z$ is multi-valued.  To make $Z$ single-valued, we choose a branch cut in which $Z_a=|Z|e^{2i\pi}$ along the anti-Stokes line labeled $a$.  Note that the anti-Stokes lines labeled $b$ and $c$ are rotated relative to $a$ by ${\frac{-2\pi}{3}}$ and ${\frac{-4\pi}{3}}$ (in the $\chi$-plane) respectively.  Therefore, we get $Z_b=|Z|e^{i\pi}$ along the anti-Stokes line labeled $b$ and $Z_c=|Z|$ along the anti-Stokes line labeled $c$ in Fig.\ \ref{two-paths}.  %The choice of branch cuts should not interfere with the path ({\it path 1} or {\it path 2} in Fig.\ \ref{two-paths}) we take to derive the WKB condition on the QNM frequency.  
We now can take the same steps as in Eqs.\ (\ref{holomorphic-bessel}) through (\ref{asymptotic-bessel}) 
%Since
%\begin{equation}
%J_{\pm \nu}(Z)=Z^{\pm \nu}\phi(Z)~,
%\end{equation}
%where $\phi$ is an even holomorphic function, we can write
%\begin{equation}
%J_{\pm\nu}\left(e^{i \theta} e^{-i \theta}Z\right)=e^{\pm i\nu\theta}J_{\pm\nu}\left(e^{-i\theta}Z\right)~,
%\end{equation}
%where $\theta=\arg{Z}$.
%Since $e^{-i\theta}Z\gg 1$ in the $|\omega| \rightarrow \infty $ limit, we can use the asymptotic behavior of the Bessel function
%\begin{equation}
%J_{\pm\nu}(e^{-i\theta}Z)\rightarrow\sqrt{\frac{2}{\pi{e^{-i\theta}Z}}}\cos\left(e^{-i\theta}Z\mp\frac{\pi \nu}{2}-\frac{\pi}{4}\right)~
%\label{asymptotic-bessel}
%\end{equation}
to show that 
\begin{eqnarray}
\psi&\approx&A_+2e^{i\frac{2\theta}{\pi}\alpha_+}\sqrt{\frac{-2GM}{\omega\sqrt{2ik}\sqrt{\chi-ik}}}\cos\left(e^{-i\theta}Z-\alpha_+\right)\nonumber \\
&+& A_-2e^{i\frac{2\theta}{\pi}\alpha_-}\sqrt{\frac{-2GM}{\omega\sqrt{2ik}\sqrt{\chi-ik}}}\cos\left(e^{-i\theta}Z-\alpha_-\right)~,
\end{eqnarray}
where $\theta=\arg{Z}$ and $\alpha_\pm =\frac{\pi}{4}\left(1\pm\sqrt{-\frac{1}{3}}\right)$.  Note that the above approximation of Bessel functions is only valid if $|Z|>>1$.  For $|Z|$ to be large, we need to stay away from the pole at $\chi=ik$ at a distance where 
\begin{equation}
|\chi-ik|>>{\left( \frac{GM}{\sqrt{k}} |\omega|\right)^{\frac{2}{3}}}~.
\end{equation}
In addition, for our calculations to be valid, the above distance should be much smaller than the polymerization length scale $k$, which determines the distance between the poles at $\pm ik$ and $0$.  This way our calculations near the pole at $ik$ will not get affected by the presence of the pole at $\chi=0$.  This leads to a lower bound on the size of $|\omega|$, i.e.
\begin{equation}
|\omega|>>{\frac{GM}{k^2}}~.
\label{lower-bound-omega}
\end{equation}
%This means that our calculations for the topology shown in Fig.\ \ref{two-paths} is valid for the range of damping rate of
%\begin{equation}
%{2GM \over k^2}<<|\omega| \lesssim {(2GM)^2 \over k^3}~.
%\end{equation}
%While this is a small range for black holes with $k\approx 2GM$, but it is a wide range for black holes in which $k<< 2GM$. 
 
For $e^{-i\theta}=+1$, we get
\begin{eqnarray}
\psi&=&   \left(A_+e^{-i\alpha_+(1-\frac{2\theta}{\pi})}+A_-e^{-i\alpha_-(1-\frac{2\theta}{\pi})}\right)f_1^{(ik)} \nonumber\\
&+& \left(A_+e^{i\alpha_+(1+\frac{2\theta}{\pi})}+A_-e^{i\alpha_-(1+\frac{2\theta}{\pi})}\right)f_2^{(ik)}~,
\label{+ik-general1}
\end{eqnarray}
and for $e^{-i\theta}=-1$, we get
\begin{eqnarray}
\psi&=&  \left(A_+e^{i\alpha_+(1+\frac{2\theta}{\pi})}+A_-e^{i\alpha_-(1+\frac{2\theta}{\pi})}\right)f_1^{(ik)} \nonumber \\
&+& \left(A_+e^{-i\alpha_+(1-\frac{2\theta}{\pi})}+A_-e^{-i\alpha_-(1-\frac{2\theta}{\pi})}\right)f_2^{(ik)}~,
\label{+ik-general2}
\end{eqnarray}
where
\begin{equation}
f_1^{(ik)}=\sqrt{\frac{-2GM}{\omega\sqrt{2ik}\sqrt{\chi-ik}}} e^{iZ}   
\end{equation}
and
\begin{equation}
f_2^{(ik)}=\sqrt{\frac{-2GM}{\omega\sqrt{2ik}\sqrt{\chi-ik}}} e^{-iZ}   
\end{equation}
are the approximate WKB solutions to the wave equation (\ref{ik-approx-wave-eq}) at a distance $>>{\left( \frac{GM}{\sqrt{k} |\omega|}\right)^{\frac{2}{3}}}$ from the pole at $ik$.  Using Eqs.\ (\ref{+ik-general1}) and (\ref{+ik-general2}) we can determine the wave solutions along the anti-Stokes lines labeled $a$, $b$, and $c$:
\begin{eqnarray}
\psi_a &=&   \left(A_+e^{3i\alpha_+}+A_-e^{3i\alpha_-}\right)f_1^{(ik)} (Z_a) 
+ \left(A_+e^{5i\alpha_+}+A_-e^{5i\alpha_-}\right)f_2^{(ik)}(Z_a)~, \nonumber \\
\psi_b &=&   \left(A_+e^{3i\alpha_+}+A_-e^{3i\alpha_-}\right)f_1^{(ik)}(Z_b)  
+ \left(A_+e^{i\alpha_+}+A_-e^{i\alpha_-}\right)f_2^{(ik)}(Z_b)~, \nonumber \\
\psi_c &=&   \left(A_+e^{-i\alpha_+}+A_-e^{-i\alpha_-}\right)f_1^{(ik)}(Z_c)
+ \left(A_+e^{i\alpha_+}+A_-e^{i\alpha_-}\right)f_2^{(ik)}(Z_c)~. \nonumber \\
\end{eqnarray}

We now move to the pole at $\chi=0$, where the QNM wave equation takes the form
\begin{equation}
\frac{d^2\psi}{d\chi^2}+\left[\frac{\omega^2k^2}{(-2GM)^2}+\frac{1}{\chi^2}\right]\psi=0~,
\label{wave-eq-0-qc}
\end{equation}
with the Bessel solution
\begin{equation}
\psi=B_+\sqrt{2\pi\chi}J_{\frac{\sqrt{-3}}{2}}\left(Z_0\right)+ B_-\sqrt{2\pi\chi}J_{-\frac{\sqrt{-3}}{2}}\left(Z_0\right)~, 
\end{equation}
where $B_{\pm}$ are constants and
\begin{equation}
Z_0=\frac{\omega k}{-2GM}\chi~.
\label{eq:Z0}
\end{equation}
To make $Z_0$ single-valued, we choose the branch cut in which $Z_{0d}=|Z_{0}|e^{i\pi}$ along the anti-Stokes line labeled $d$.  A $-\pi$ rotation in the $\chi$-plane will take us to the line labeled $e$ on which $Z_{0e}=Z_{0d}e^{-i\pi}=|Z_{0}|$.  Once again, we can use the asymptotic behavior of Bessel functions (\ref{asymptotic-bessel}) in the large $Z_0$ limit to show that 
\begin{eqnarray}
\psi&=&B_+2e^{i\frac{2\theta_0}{\pi}\beta_+}\sqrt{\frac{-2GM}{\omega k}} \cos\left(e^{-i\theta_0}Z_0-\beta_+\right)\nonumber \\
&+& B_-2e^{i\frac{2\theta_0}{\pi}\beta_-}\sqrt{\frac{-2GM}{\omega k}}\cos\left(e^{-i\theta_0}Z_0-\beta_-\right)~,
\end{eqnarray}
where $\theta_0=\arg Z_0$ and $\beta_\pm =\frac{\pi}{4}\left(1\pm\sqrt{-3}\right)$.   From Eq.\ (\ref{eq:Z0}), it is clear that the condition, which allows $Z_0$ to be large without getting too close to the other poles, is precisely that given in Eq.\ (\ref{lower-bound-omega}).  This is required for the validity of the calculation. 

For $e^{-i\theta_0}=+1$, we get
\begin{eqnarray}
\psi&=&   \left(B_+e^{-i\beta_+(1-\frac{2\theta_0}{\pi})}+B_-e^{-i\beta_-(1-\frac{2\theta_0}{\pi})}\right)f_1^{(0)} \nonumber\\
&+&  \left(B_+e^{i\beta_+(1+\frac{2\theta_0}{\pi})}+B_-e^{i\beta_-(1+\frac{2\theta_0}{\pi})}\right)f_2^{(0)}~,
\end{eqnarray}
and for $e^{-i\theta}=-1$, we get
\begin{eqnarray}
\psi&=&  \left(B_+e^{i\beta_+(1+\frac{2\theta_0}{\pi})}+B_-e^{i\beta_-(1+\frac{2\theta_0}{\pi})}\right)f_1^{(0)} \nonumber \\
&+& \left(B_+e^{-i\beta_+(1-\frac{2\theta_0}{\pi})}+B_-e^{-i\beta_-(1-\frac{2\theta_0}{\pi})}\right)f_2^{(0)}~,
\end{eqnarray}
where
\begin{equation}
f_1^{(0)}=\sqrt{\frac{-2GM}{\omega k}} e^{iZ_0}   
\end{equation}
and
\begin{equation}
f_2^{(0)}=\sqrt{\frac{-2GM}{\omega k}} e^{-iZ_0}   
\end{equation}
are the approximate WKB solutions of the wave equation (\ref{wave-eq-0-qc}) when $|\chi| >>   {\frac{GM}{k |\omega|} }$.  Using these equations we can determine the wave solutions along the anti-Stokes lines labeled $d$ and $e$ to be:
\begin{eqnarray}
\psi_d &=&   \left(B_+e^{3i\beta_+}+B_-e^{3i\beta_-}\right)f_1^{(0)} (Z_{0d}) 
+ \left(B_+e^{i\beta_+}+B_-e^{i\beta_-}\right)f_2^{(0)}(Z_{0d})~, \nonumber \\
\psi_e &=&   \left(B_+e^{-i\beta_+}+B_-e^{-i\beta_-}\right)f_1^{(0)}(Z_{0e})
+ \left(B_+e^{i\beta_+}+B_-e^{i\beta_-}\right)f_2^{(0)}(Z_{0e})~. \nonumber \\
\label{psi-d-e}
\end{eqnarray}

We finally move close to $\chi=-ik$, where the differential equation (\ref{de-qnm-pk}) can be expressed as
\begin{equation}
\frac{d^2\psi}{d\chi^2}+\left[\frac{\omega^2(-2ik)(\chi+ik)}{(-2GM)^2}+\frac{7}{16(\chi+ik)^2}\right]\psi=0
\label{-ik-wave-eq}
\end{equation}
with the solution 
\begin{equation}
\psi=C_+\sqrt{\frac{4\pi}{3}(\chi+ik)}J_{\nu}\left(\bar{Z}\right)+ C_-\sqrt{\frac{4\pi}{3}(\chi+ik)}J_{-\nu}(\bar{Z})~,
\end{equation}
where  $C_\pm$ are constants and
\begin{equation}
\bar{Z}=\frac{2}{3}\left(\frac{\omega\sqrt{-2ik}}{-2GM}\right)(\chi+ik)^{3/2}~.
\end{equation}
To make $\bar{Z}$ single-valued, we choose a branch cut in which $\bar{Z}_f=|\bar{Z}|e^{i\pi}$ along the anti-Stokes line labeled $f$, $\bar{Z}_g=|\bar{Z}|$ along the anti-Stokes line labeled $g$ and $\bar{Z}_h=|\bar{Z}|e^{-i\pi}$ along the anti-Stokes line labeled $h$.   
Following similar steps as in previous cases, we find 
\begin{eqnarray}
\psi_f &=&   \left(C_+e^{3i\alpha_+}+C_-e^{3i\alpha_-}\right)f_1^{(-ik)}(\bar{Z}_f)  
+ \left(C_+e^{i\alpha_+}+C_-e^{i\alpha_-}\right)f_2^{(-ik)}(\bar{Z}_f)~, \nonumber \\
\psi_g &=&   \left(C_+e^{-i\alpha_+}+C_-e^{-i\alpha_-}\right)f_1^{(-ik)}(\bar{Z}_g) 
+ \left(C_+e^{i\alpha_+}+C_-e^{i\alpha_-}\right)f_2^{(-ik)}(\bar{Z}_g)~, \nonumber \\
\psi_h &=&   \left(C_+e^{-i\alpha_+}+C_-e^{-i\alpha_-}\right)f_1^{(-ik)}(\bar{Z}_h)  
+  \left(C_+e^{-3i\alpha_+}+C_-e^{-3i\alpha_-}\right)f_2^{(-ik)}(\bar{Z}_h)~, \nonumber \\
\end{eqnarray}
where
\begin{equation}
f_1^{(-ik)}=\sqrt{\frac{-2GM}{\omega\sqrt{-2ik}\sqrt{\chi+ik}}} e^{i\bar{Z}}   
\end{equation}
and
\begin{equation}
f_2^{(-ik)}=\sqrt{\frac{-2GM}{\omega\sqrt{-2ik}\sqrt{\chi+ik}}} e^{-i\bar{Z}}   
\end{equation}
are the approximate WKB solutions to the wave equation (\ref{-ik-wave-eq}) at a distance $>>{\left( \frac{GM}{\sqrt{k} |\omega|}\right)^{\frac{2}{3}}}$ from the pole at $-ik$.

The solutions do not change in character along anti-Stokes lines.  This means that, since the anti-Stokes line labeled $a$ extends to infinity, we can impose the boundary condition at infinity on $\psi_a$.  This gives us the condition   
\begin{equation}
A_+e^{5i\alpha_+}+A_-e^{5i\alpha_-}=0~.
\label{e1}
\end{equation}
Also, $\psi_c$ and $\psi_d$, which are on the same anti-Stokes line, have to be equal.  Since we can write  
\begin{equation}
\psi_c =  \left(A_+e^{-i\alpha_+}+A_-e^{-i\alpha_-}\right)e^{\delta_{cd}}f_1^{(0)}(Z_d)  
+ \left(A_+e^{i\alpha_+}+A_-e^{i\alpha_-}\right)e^{-\delta_{cd}}f_2^{(0)}(Z_d)~,
\label{psi_c}
\end{equation}
where
\begin{equation}
\delta_{cd}=\int_{ik}^0 \tilde{Q} d\chi \approx \int_{ik}^0 \left(\frac{\omega \sqrt{\chi^2+k^2}}{\chi-2GM}\right)d\chi \approx \int_{ik}^0 \left(\frac{\omega\sqrt{\chi^2+k^2}}{-2GM}\right) d\chi = \frac{i\pi \omega  k^2}{8GM}~
\end{equation} 
appears as a result of changing the lower limit of the phase integral of the WKB solutions from $ik$ to $0$.  We can now compare (\ref{psi_c}) with $\psi_d$ given in Eq.\ (\ref{psi-d-e}), which gives the following two conditions: 
\begin{equation}
\left(A_+e^{-i\alpha_+}+A_-e^{-i\alpha_-}\right)e^{i\delta_{cd}}=B_+e^{3i\beta_+}+B_-e^{3i\beta_-}~,
\label{e2}
\end{equation} 
\begin{equation}
\left(A_+e^{i\alpha_+}+A_-e^{i\alpha_-}\right)e^{-i\delta_{cd}}=B_+e^{i\beta_+}+B_-e^{i\beta_-}~.
\label{e3}
\end{equation}  
The same thing can be done for $\psi_e$ and $\psi_f$, which results in two more conditions:
\begin{equation}
\left(B_+e^{-i\beta_+}+B_-e^{-i\beta_-}\right)e^{i\delta_{ef}}=C_+e^{3i\alpha_+}+C_-e^{3i\alpha_-}~,
\label{e4}
\end{equation}   
\begin{equation}
\left(B_+e^{i\beta_+}+B_-e^{i\beta_-}\right)e^{-i\delta_{ef}}=C_+e^{i\alpha_+}+C_-e^{i\alpha_-}~,
\label{e5}
\end{equation} 
where 
\begin{equation}
\delta_{ef}=\int_{0}^{-ik} \tilde{Q} d\chi=  \delta_{cd}.
\end{equation} 
So far, all the conditions above apply to both {\it path 1} and {\it path 2} in Fig.\ (\ref{two-paths}).     
In the case of {\it path 1}, since the anti-Stokes line $h$ extends to infinity, we can impose the boundary condition at infinity to $\psi_h$, which gives us the condition:
\begin{equation}
C_+e^{-3i\alpha_+}+C_-e^{-3i\alpha_-}=0~.
\end{equation} 
This condition however does not impose any WKB condition on the QNM frequency.  Another condition, that we can get, is by moving from point $h$ to infinity along the anti-Stokes line that extends to infinity in the lower half of the complex $\chi$-plane.  Then we can rotate at infinity, where we know the solution due to the boundary condition, to the other anti-Stokes line which extends to infinity in the upper half of the complex plane.  Finally we can move along this anti-Stokes line all the way back to point $a$, where we get 
\begin{equation}
\psi_{h}=\left(C_+e^{-i\alpha_+}+C_-e^{-i\alpha_-}\right)e^{i\tilde{\delta}_{ha}}f_1^{(ik)}(Z_a)+\mbox{sub-dominant term}=\psi_{\bar{a}}~. 
\end{equation} 
For this path the coefficient of the sub-dominant solution $f_2$ is not reliable.
Here
\begin{equation}
\tilde{\delta}_{ha}= \int_{-ik}^{ik} \tilde{Q} d\chi 
\end{equation} 
along the path to the right of the event horizon.  By returning back to point $a$, we have completed a loop around the event horizon.  Therefore, in matching $\psi_a$ and $\psi_{\bar{a}}$ we need to account for the monodromy of {\it path 1}.  In other words  
\begin{equation}
\psi_{\bar{a}}=e^{-i\Delta_{ccw}}\psi_a~,
\label{final-condition}
\end{equation}
where $\Delta_{ccw}$ is the integral of $\tilde{Q}$ along a contour encircling the pole at $\chi=2GM$ in the counter-clockwise ($ccw$) direction:
\begin{equation}
\Delta_{ccw}=\mathop{\oint}_{ccw} \tilde{Q} d\chi \approx 2\pi i  \mathop{Res}_{\chi=2GM}\left(\frac{\omega \sqrt{\chi^2+k^2}}{\chi-2GM}\right)=2\pi i \omega\sqrt{4G^2M^2+k^2}~.    
\end{equation} 
Equation (\ref{final-condition}) gives us the final condition
\begin{equation}
\left(C_+e^{-i\alpha_+}+C_-e^{-i\alpha_-}\right)e^{i\Delta_{ccw}}e^{-i(\delta_{cd}+\delta_{ef})}=\left(A_+e^{3i\alpha_+}+A_-e^{3i\alpha_-}\right)e^{-i\Delta_{ccw}}~,
\label{e6}
\end{equation} 
where we have replaced $\tilde{\delta}_{ha}$ with $\Delta_{ccw}-\delta_{cd}-\delta_{ef}$.

In the case of {\it path 2}, we first need to match the solution at $g$ with the solution at $b$, which are on the same anti-Stokes line.  This matching results in two conditions:
\begin{equation}
\left(C_+e^{-i\alpha_+}+C_-e^{-i\alpha_-}\right)e^{i\Delta_{ccw}}e^{-i(\delta_{cd}+\delta_{ef})}=\bar{A}_+e^{3i\alpha_+}+\bar{A}_-e^{3i\alpha_-}~,
\label{path2-e1}
\end{equation}    
\begin{equation}
\left(C_+e^{i\alpha_+}+C_-e^{i\alpha_-}\right)e^{-i\Delta_{ccw}}e^{i(\delta_{cd}+\delta_{ef})}=\bar{A}_+e^{i\alpha_+}+\bar{A}_-e^{i\alpha_-}~. 
\label{path2-e2}
\end{equation}  
Here we have used the fact that
\begin{equation}
\tilde{\delta}_{gb}=\int_{-ik}^{ik} \tilde{Q} d\chi= \Delta_{ccw}-\delta_{cd}-\delta_{ef}~,
\end{equation} 
where $\tilde{\delta}_{gb}$ is taken along the anti-Stokes line to the right of the pole at the event horizon.   
We finally can return to anti-Stokes line $a$ where we can impose the boundary condition at infinity to the coefficient of the sub dominant solution to get
\begin{equation}
\bar{A}_+e^{5i\alpha_+}+\bar{A}_-e^{5i\alpha_-}=0~.
\label{path2-e3}
\end{equation}
We can also impose the monodromy condition that $\psi_{\bar{a}}=e^{-i\Delta_{ccw}}\psi_a$ on the coefficient of the dominant solution to find 
\begin{equation}
\bar{A}_+e^{3i\alpha_+}+\bar{A}_-e^{3i\alpha_-}=\left(A_+e^{3i\alpha_+}+A_-e^{3i\alpha_-}\right)e^{-i\Delta_{ccw}}~.
\label{path2-e3'}
\end{equation}
%This condition will result in the same WKB condition on the QNM frequency as the one in Eq.\ (\ref{path2-e3}).  

Finally, we can use Eqs.\ (\ref{e1}), (\ref{e2}), (\ref{e3}), (\ref{e4}) and (\ref{e5}) together with either (\ref{e6}) from {\it path 1} or with the combination of Eqs.\ (\ref{path2-e1}), (\ref{path2-e2}) and  (\ref{path2-e3}) or (\ref{path2-e3'}) from {\it path 2} to get a WKB condition on $\omega$.  All the above combinations should result in the same WKB condition on the QNM frequency, which they do.  This provides a valuable check on the validity of our results.  For the case of {\it path 1}, in order to have a nontrivial solution to Eqs.\ (\ref{e1}), (\ref{e2}), (\ref{e3}), (\ref{e4}), (\ref{e5}) and (\ref{e6}), the determinant of the matrix	 
%%%%%%%%%%%%%%%%%%%%%%%%%%%%%%%%%%%%%%%%%%%%%%%%%\[
\begin{eqnarray}
\left[ 
{\begin{array}{cccccc}
e^{5i\,\alpha_+} & e^{5i\,\alpha_-} & 0 & 0 & 0 & 0 \\
e^{-i\,\alpha_+}e^{i\delta} & e^{-i\,\alpha_-}\,e^{i\delta} &  - e^{3i\,
\beta_+} &  - e^{3i\,\beta_-} & 0 & 0 \\
{\displaystyle  {e^{i\,\alpha_+}}e^{-i\delta}}  & 
{\displaystyle {e^{i\,\alpha _-}e^{-i\delta}}}  &  - e^{i\,
\beta_+} &  - e^{i\,\beta_-} & 0 & 0 \\ [2ex]
0 & 0 & e^{ -i\,\beta_+}e^{i\delta} & e^{-i\,\beta_-}e^{i\delta} &  - 
{\displaystyle {e^{3i\,\alpha_+}}}  &  - 
{\displaystyle {e^{3i\,\alpha_-}}}  \\ [2ex]
0 & 0 & e^{i\beta_+}e^{-i\delta} & e^{i\beta_-}e^{-i\delta} &  - e^{i\,
\alpha_+} &  - e^{i\,\alpha_-} \\
e^{3i\alpha_+} & e^{3i\alpha_-} & 0 & 0 &  - 
{\displaystyle  {e^{-i\alpha_+}}e^{2i\Delta_{ccw}}e^{-2i\delta}}  &  - 
{\displaystyle {e^{-i\alpha_-}}e^{2i\Delta_{ccw}}e^{-2i\delta}}  
\end{array}}
 \right]\nonumber\\
\end{eqnarray}
%%%%%%%%%%%%%%%%%%%%%%%%%%%%%%%%%%%%%%%%%%%%%%%%%
has to vanish.  This leads to the WKB condition
\begin{eqnarray}
e^{-2i\Delta_{ccw}}+K_1e^{-4i\delta}+  
K_2e^{-2i\delta} +K_3=0~,
\end{eqnarray}
where
\begin{eqnarray}
\delta &=& \delta_{cd}=\delta_{ef},\nonumber\\
K_1&=&4\cos^2(\alpha_+-\alpha_-)~, \nonumber\\
K_2&=&4\cos(\alpha_+-\alpha_-)\cos(\beta_+-\beta_-)~,\nonumber\\
K_3&=&4\cos^2(\alpha_+-\alpha_-)-1~.
\end{eqnarray}
Since $\Delta_{ccw}\approx4\pi i \omega GM+4\delta$, the above equation can be written as
\begin{eqnarray}
e^{8\pi\omega GM}+K_1e^{-\frac{\pi\omega k^2}{2GM}}+  
K_2e^{-\frac{3\pi\omega k^2}{4GM}}+K_3e^{-\frac{\pi\omega k^2}{GM}}=0~.
\end{eqnarray}
We now take $y=8\pi\omega GM$ to get
\begin{eqnarray}
e^{y}+K_1e^{-2gy}+K_2e^{-3gy}+K_3e^{-4gy}=0
\end{eqnarray}
where $g=\frac{k^2}{32G^2M^2}$.
We then take $y=\zeta-i\eta$, which leads to
\begin{eqnarray}
e^{\zeta}\cos \eta+K_1e^{-2g\zeta}\cos(2g\eta)+  
K_2e^{-3g\zeta}\cos(3g\eta)+K_3e^{-4g\zeta}\cos(4g\eta)=0~
\end{eqnarray}
and
\begin{eqnarray}
-e^{\zeta}\sin \eta+K_1e^{-2g\zeta}\sin(2g\eta)+  
K_2e^{-3g\zeta}\sin(3g\eta)+K_3e^{-4g\zeta}\sin(4g\eta)=0~.
\end{eqnarray} 
From the last two equations, we see that for the solution to be periodic in $\eta$, we must require simultaneously
\begin{eqnarray}
\eta\rightarrow \eta+2n\pi ~~\mbox{and}~~ g(\eta+2n\pi)=g\eta+2m\pi~,
\end{eqnarray} 
where $n$ and $m$ are integers.  In other words, periodicity in the damping term is only possible if
\begin{eqnarray}
g=\frac{m}{n}~.
\end{eqnarray} 
Note that since $g=\frac{k^2}{32G^2M^2}$, we require $\frac{m}{n}\ll 1$.

\begin{figure}[tb]
\begin{center}
\includegraphics[height=3.cm]{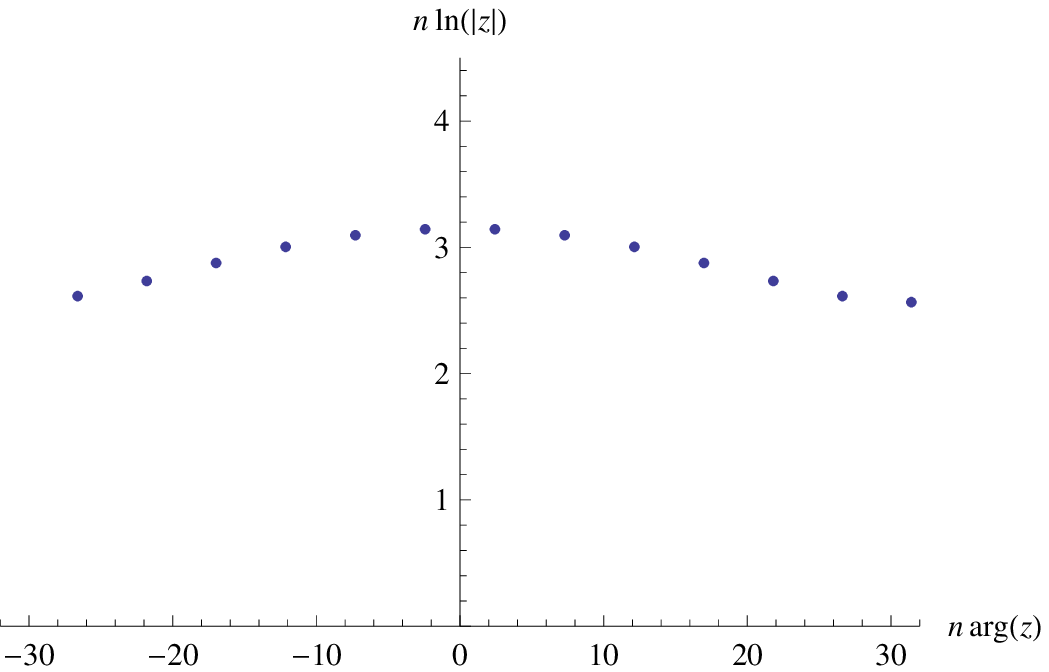}
\includegraphics[height=3.cm]{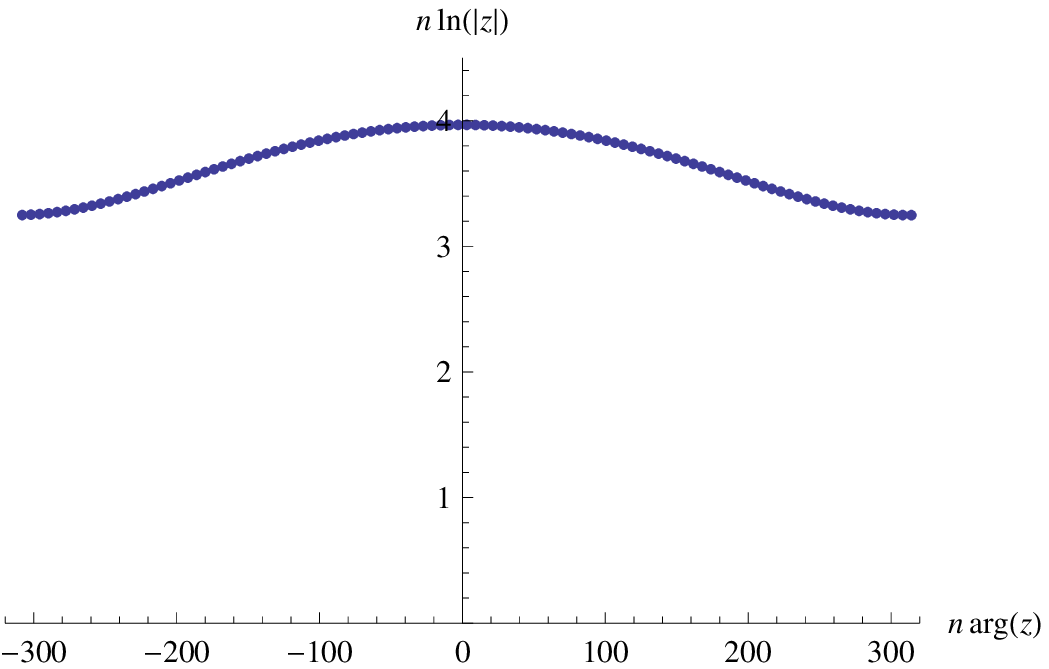}
\includegraphics[height=3.cm]{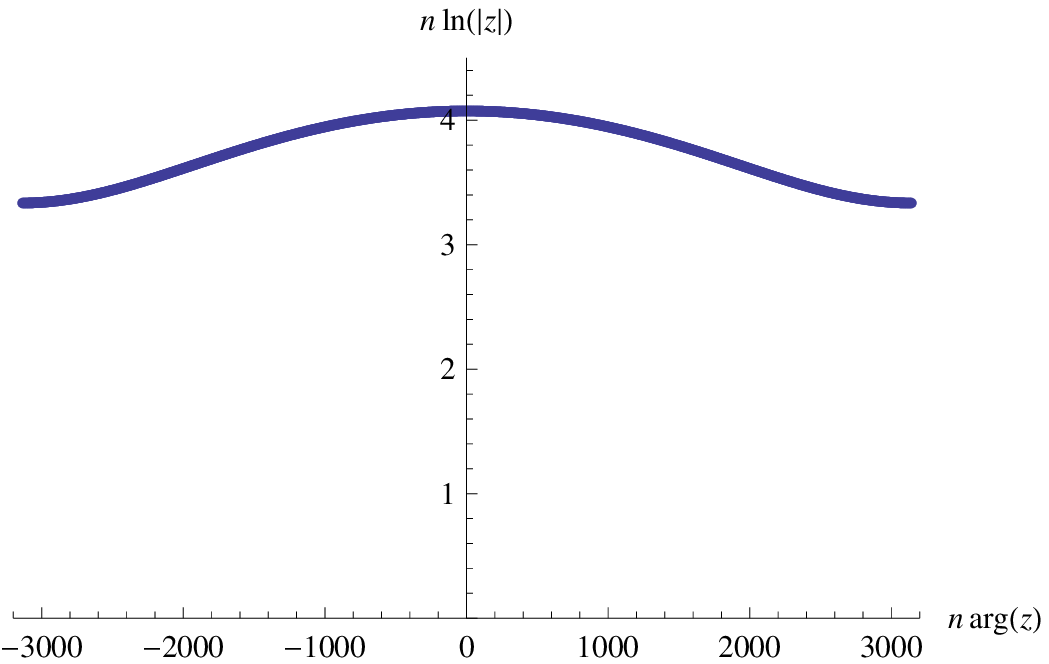}
\end{center}
\caption{\footnotesize Real part of $8\pi\omega GM$ versus the imaginary part for $m=1$ and $n=10,~100,~1000$ from left to right respectively.}
\label{finite-k-numeric}
\end{figure}

For $m$ and $n$ integers we can introduce $z=e^{y/n}$ to obtain
\begin{eqnarray}
z^{n+4m}+K_1z^{2m}+K_2z^m+K_3=0~,
\label{roots-PK}
\end{eqnarray} 
which has $n+4m$ roots.  (Note that in $z^{n+4m}$ above, one can neglect $4m$ since $n>>m$.  This is equivalent to neglecting the term $4\delta$ in $\Delta_{ccw}$.)  These roots lead to
\begin{equation}
8\pi \omega GM=y=n\ln|z|+in\arg (z)-2np\pi i~,
\end{equation} 
where $p$ is a large integer.  Using the above equation, we have plotted the QNM frequency spectrum in Fig.\ \ref{finite-k-numeric} for $m=1$ and $n=10,~100,~1000$.    

In the limit $k\rightarrow 0$, Eq.\ (\ref{roots-PK}) leads us to
\begin{equation}
e^{8\pi \omega GM}=-(K_1+K_2+K_3)=-59.5392...~.
\end{equation} 
This gives a QNM frequency spectrum of the form
\begin{equation}
8\pi \omega GM=\ln(59.5392...)-2 \pi i \left(l+{\frac{1}{2}}\right)~,
\end{equation} 
where $l$ is a large integer.  This is different than the Schwarzschild result due to the presence of the quantum length scale $k$ in the spacetime metric.

We, finally, would like to point out that in the intermediate damping region of (\ref{intermediate-w-range}), our calculations in the complex plane are done at a distance of $|\chi| >> k$ from the origin of the complex plane.  This is the reason why we get the Schwarzschild value of $\ln(3)$ for the real part of $\omega$.  In this intermediate damping region, our path will not be affected by the ``fine structure" of the poles near $\chi=0$ because, for example in {\it path 1}, the jump from anti-Stokes line $a$ to line $h$  in Fig.\ \ref{two-paths} is done at a distance scale much larger than $k$.

\section{Conclusions}
We have presented an analytic calculation of the highly damped QNM's of a non-singular QC black hole using the methods of Motl and Neitzke. As expected, the spectrum of highly damped QNM's are sensitive to the metric structure at the new scale,  $k$, which determines the onset of the quantum corrections. In analogy with what happens in the R-N case, the real part of the highly damped QNM frequency does not reduce to the Schwarzschild value even in the limit that $k\to0$. As in the R-N case, this is explained by the fact that the limit $|\omega|
\to\infty$ does not commute with the limit $k\to 0$.  The present calculation suggests that other non-singular QC metrics with different small scale behavior will yield a different answer in this limit. This is currently under investigation. The implication is that the highly damped QNM's in principle, if not in practice, contain information about the underlying theory that is used to resolve the singularity.

Our analysis is also interesting for the following reason. We found that, in analogy with the Gauss-Bonnet black hole\cite{dkz06}, the AH method was difficult to implement with the given topology of Stokes and anti-Stokes lines of the QC black hole. This is related to the fact that both metrics have a branch point singularity.  While these difficulties with the AH method are likely surmountable, we have shown that the MN method does in fact appear to produce an unambiguous answer in the present case in a rather straightforward fashion. This suggests that it may be possible to get a consistent solution for the more complicated Gauss-Bonnet black hole using the MN method. This calculation is also under investigation.

Finally, we stress again the fact that the analytically calculated high overtone QNM's probe the structure of the black hole exterior down to length scales determined by the inverse of the magnitude of the QNM frequency.  Whatever corrections to the small scale structure result from the ultimate quantum gravity theory, QNM's which are large relative to the inverse horizon length but small compared to other inverse length squares will be insensitive to these corrections and can in principle provide information only about the horizon structure. 

%Finally, the fact that the analytically calculated high overtone QNM's probe the structure of the black hole solution to length scales determined by the inverse of the magnitude of the QNM frequency tells us that, whatever the ultimate quantum corrected solution does to the small scale analytic structure, it is plausible to think that the QNM's that are large relative to the inverse horizon length but small compared to other inverse length squares will provide information about the event horizon.   

\section{Acknowledgments} We are grateful to Eric Poisson for useful conversations.  We are also thankful to Michael Green for sharing his Mathematica files with us to generate the anti-Stokes line structures numerically.  This work was supported in part by the Natural Science and Engineering Research Council of Canada.
%%%%%%%%%%%%%%%%%%%%%%%%%%%%%%

% A useful Journal macro
\def\jnl#1#2#3#4{{#1}{\bf #2}, #3 (#4)}

\def\Zphys{{Z.\ Phys.} }
\def\jssc{{J.\ Solid State Chem.\ }}
\def\jpsJ{{J.\ Phys.\ Soc.\ Japan }}
\def\ptps{{Prog.\ Theoret.\ Phys.\ Suppl.\ }}
\def\PTP{{Prog.\ Theoret.\ Phys.\  }}
\def\LNC{{Lett.\ Nuovo.\ Cim.\  }}

\def\JMP{{J. Math.\ Phys.} }
\def\NPB{{Nucl.\ Phys.} B}
\def\NP{{Nucl.\ Phys.} }
\def\PLB{{Phys.\ Lett.} B}
\def\PL{{Phys.\ Lett.} }
\def\PRL{Phys.\ Rev.\ Lett.~}
\def\PRB{{Phys.\ Rev.} B}
\def\PRD{{Phys.\ Rev.} D }
\def\PR{{Phys.\ Rev.} }
\def\PRe{{Phys.\ Rep.} }
\def\AP{{Ann.\ Phys.\ (N.Y.)} }
\def\RMP{{Rev.\ Mod.\ Phys.} }
\def\ZPC{{Z.\ Phys.} C}
\def\SCI{Science}
\def\CMP{Comm.\ Math.\ Phys. }
\def\MPLA{{Mod.\ Phys.\ Lett.} A}
\def\IJMPB{{Int.\ J.\ Mod.\ Phys.} B}
\def\cmp{{Com.\ Math.\ Phys.}}
\def\JPA{{J.\  Phys.} A}
\def\CQG{Class.\ Quant.\ Grav.~}
\def\ATMP{Adv.\ Theoret.\ Math.\ Phys.~}
\def\PRSA{{Proc.\ Roy.\ Soc.} A }
\def\ibid{{ibid.} }

\end{document}